\documentclass[preprint,12pt]{elsarticle}

\usepackage{amssymb}
\usepackage{amsmath}
\usepackage{siunitx}
\usepackage{graphicx}
\usepackage{subcaption}
\usepackage{hyperref}
\usepackage{booktabs}
\usepackage{multirow}
\usepackage{makecell}
\usepackage{tikz}
\usepackage{hyphenat}
\usepackage{here}

\usepackage{lineno}

\journal{Nucl. Instrum. Methods A}

\begin{document}

\begin{frontmatter}



\title{Design and Performance of a Monolithic Plastic Scintillator Tracker with Embedded Scatterers}


\author[a]{Naoki~Otani\corref{cor1}} 
\ead{otani.naoki.53x@st.kyoto-u.ac.jp}
\author[a]{Seungho~Han}
\author[b]{Shun~Ito\fnref{fn1}}
\author[a]{Tatsuya~Kikawa}
\author[a]{Tsuyoshi~Nakaya}
\author[b]{Mihiro~Suzuki\fnref{fn1}}
\author[c]{Atsushi Tokiyasu}
\affiliation[a]{organization={Department of Physics, Kyoto University},
            addressline={Kitashirakawa Oiwake-cho, Sakyo-ku},
            city={Kyoto},
            postcode={606-8502},
            country={Japan}}
\affiliation[b]{organization={
Department of Physics, Yokohama National University},
            addressline={79-5 Tokiwadai, Hodogaya-ku},
            city={Yokohama, Kanagawa},
            postcode={240-8501},
            country={Japan}}
\affiliation[c]{organization={Research Center for Accelerator and Radioisotope Science, Tohoku University},
            addressline={1-2-1 Mikamine, Taihaku-ku},
            city={Sendai, Miyagi},
            postcode={982-0826},
            country={Japan}}

\cortext[cor1]{Corresponding author.}
\fntext[fn1]{Currently in industry.}

\begin{abstract}
We propose a new scintillator-based tracker concept based on a monolithic plastic scintillator plate with embedded scatterers and wavelength-shifting fiber readout.
The embedded scatterers localize scintillation light so that channels closer to the charged-particle crossing point collect more light.
The particle crossing position is reconstructed from the channel-to-channel light yield distribution with a position resolution well below the readout pitch.
We performed a positron beam test with prototypes to validate the reconstruction principle and to evaluate the detection efficiency and position resolution.
The beam test validated the position reconstruction principle, and demonstrated a near-100$\%$ detection efficiency and a position resolution of $1.47~\mathrm{mm}$ for normal incidence and $1.85~\mathrm{mm}$ for an incidence angle of $45^\circ$, with a $10$-mm readout pitch.
In this paper, we describe the detector concept, the reconstruction method, and the results of the beam test.
\end{abstract}


\end{frontmatter}




\section{Introduction}
\label{sec:intro}

\begin{figure}[htbp]
  \centering
  \includegraphics[width=0.5\linewidth]{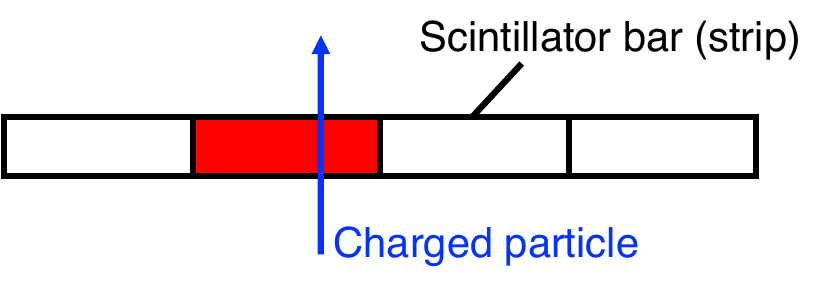}
  \caption{Schematic of a conventional segmented scintillator tracker. The fired scintillator segment is shown in red.}
  \label{fig:intro_schematics}
\end{figure}

Plastic scintillator trackers have been widely used for charged particle detection, offering fast response, simple detector construction, and scalable large-area coverage.
In many experiments, they are implemented as large planes built from segmented scintillator bars or strips, with light collected by wavelength-shifting (WLS) fibers and read out by photomultiplier tubes (PMTs) or silicon photomultipliers (SiPMs).
A representative example is the T2K ND280 fine-grained detectors~\cite{FGD}.
In such segmented designs (Fig.~\ref{fig:intro_schematics}), the particle position is identified by which bar (or strip) fired, so the achievable position resolution is fundamentally limited by the segment pitch.
In addition, segmented trackers can suffer from detection inefficiency due to inter-bar gaps and reflective coatings.

Improving the position resolution in such segmented trackers generally requires a finer segmentation pitch.
For large-area detectors, however, reducing the pitch increases the number of readout channels and consequently the cost and system complexity.
This motivates the development of tracker concepts that achieve a position resolution well below the readout pitch so that the number of channels need not be increased as the detector area increases, thereby enabling scalable and cost-effective large-area instrumentation.

A possible way to overcome the position-resolution limit imposed by segmentation is to reconstruct the hit position from the distribution of signals over multiple readout channels.
This concept has already been used in other types of detectors, such as silicon microstrip detectors~\cite{microstrip}.
In plastic scintillator detectors with WLS fiber readout, a recent study demonstrated two-dimensional position reconstruction using a fiber-structured scintillator plate with embedded WLS fibers arranged in two perpendicular directions~\cite{Brignoli2025}.
That study reported a position resolution of 5--6~mm for a WLS fiber pitch of 15~mm.
However, the reported position resolution was limited by the broad light yield distribution and by position-dependent variations in the detector response.
This suggests that localizing scintillation light near the particle crossing point could mitigate these limitations and improve the position resolution.

In this paper, we propose a new scintillator-based tracker, FROST (Fiber-Readout mOnolithic and Scatterer\hyp embedded scintillator Tracker), which consists of a monolithic plastic scintillator plate with embedded scatterers and WLS fibers embedded in grooves on the scintillator surfaces.
Scintillation light is collected by the fibers, and the particle crossing position is reconstructed from the channel-to-channel light yield distribution.
The scatterers are intentionally introduced to suppress the lateral spread of scintillation light and confine it near the particle crossing point, thereby enabling a position resolution well below the readout pitch.
Because the sensitive volume is monolithic and fully active, FROST can maintain near-100\% detection efficiency in principle.

Since the position reconstruction uses the channel-to-channel light yield distribution, nearby multiple-particle hits can lead to overlapping signal distributions and degrade the position resolution.
In this paper, we focus on the single-particle case, with applications such as low-energy neutrino experiments in mind, where the probability of multiple nearby hits is low.

The remainder of this paper is organized as follows.
Section~\ref{sec:design} describes the detector design and concept.
The Geant4-based optical simulation is presented in Sec.~\ref{sec:simulation}.
Section~\ref{sec:posrec} describes the position reconstruction method and presents simulation studies of the expected performance.
Section~\ref{sec:beamtest} reports a positron beam test performed with prototype detectors at the Research Center for Accelerator and Radioisotope Science (RARiS), Tohoku University and presents the resulting performance evaluation.
Finally, Sec.~\ref{sec:conclusion} concludes the paper.


\section{Detector design and concept}
\label{sec:design}
\begin{figure}[hbtp]
    \centering
    \begin{minipage}{0.48\textwidth}
        \centering
        \includegraphics[width=\linewidth]{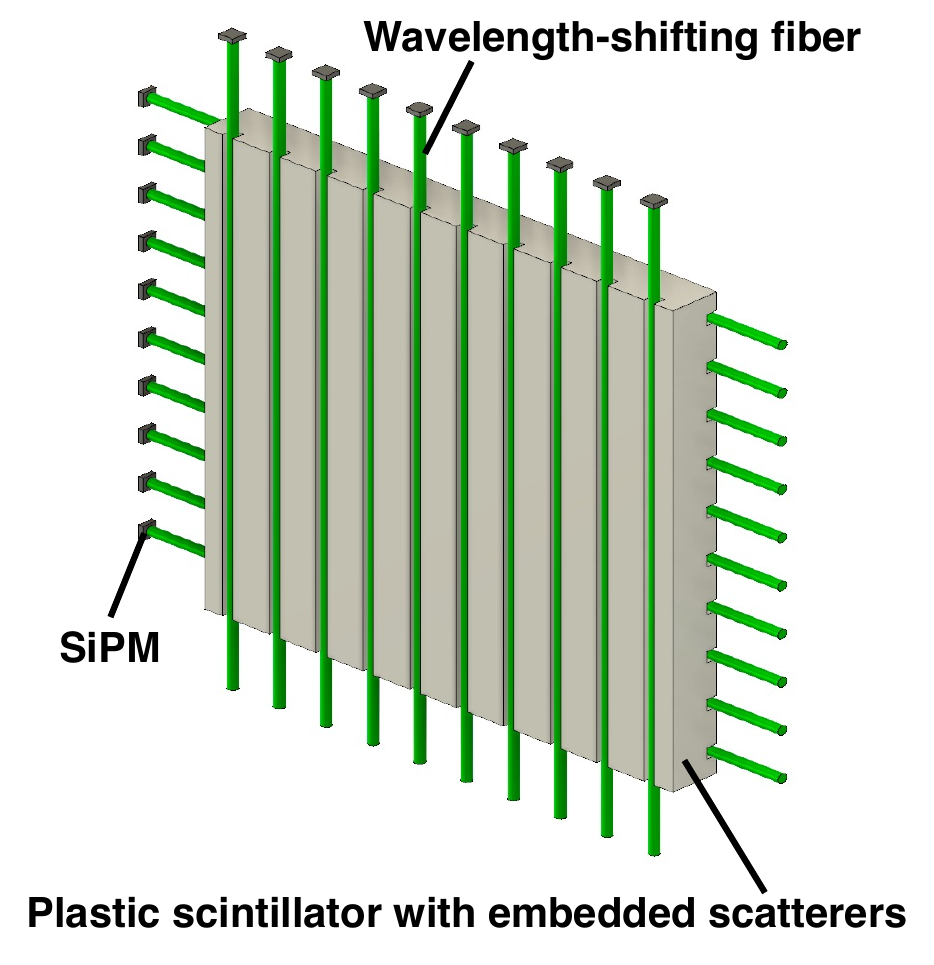}
        \caption{Conceptual design of FROST.}
        \label{fig:trackerdesign}
    \end{minipage}
    \hfill
    \begin{minipage}{0.48\textwidth}
        \centering
        \includegraphics[width=\linewidth]{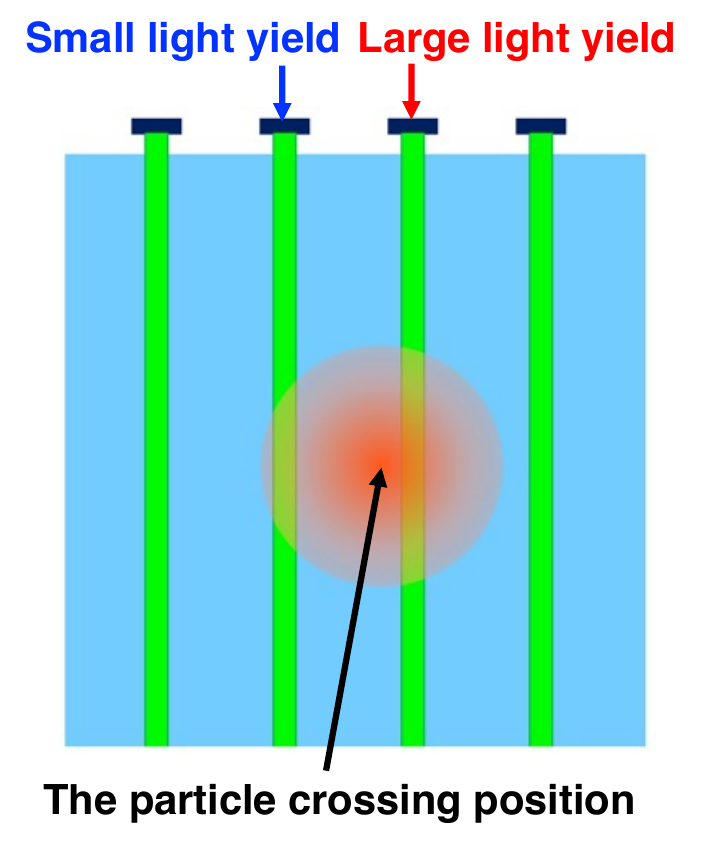}
        \caption{Mechanism of position reconstruction in FROST.}
        \label{fig:trackermechanism}
    \end{minipage}
\end{figure}

Figure~\ref{fig:trackerdesign} shows the conceptual design of FROST.
FROST consists of a monolithic plastic scintillator plate with embedded scatterers and WLS fibers placed in evenly spaced grooves on the scintillator surfaces.
The fibers are embedded on the two opposite surfaces in orthogonal directions to provide two-dimensional readout.
When a charged particle traverses the scintillator, the emitted scintillation light is collected by the fibers.
Each fiber is optically coupled to the scintillator and read out by a SiPM coupled to one end of the fiber.
The scintillator surfaces are coated with a reflective paint to enhance light collection.

As schematically shown in Fig.~\ref{fig:trackermechanism}, the collected light yield tends to be larger in channels whose fibers are closer to the particle crossing position.
The position of a charged particle passing through FROST is reconstructed based on the light yield distribution across channels.

If the scintillation light spreads over a large region, the differences in collected light yield among channels become small, degrading the position resolution.
In addition, if the detected light yield is too small, its statistical fluctuations also degrade the position resolution.
A position resolution well below the readout pitch therefore requires both sufficient localization of the scintillation light and sufficient detected light yield.
The embedded scatterers enhance the scattering of optical photons, causing them to undergo multiple scatterings within a limited region and thereby localizing the scintillation light.

\section{Optical simulation}
\label{sec:simulation}

A Geant4-based~\cite{geant4:2003,geant4:2006,geant4:2016} Monte Carlo simulation was developed to model the optical response of FROST.
The detector geometry and materials described in Sec.~\ref{sec:design} were implemented in the simulation.
Geant4 provides an optical-photon simulation framework that enables modeling of photon transport in matter, including scattering and absorption, as well as reflection and refraction at material boundaries.
Within this framework, scintillation photons were tracked in the scintillator and their interactions with the optical components were simulated, including surface reflections, wavelength shifting and light transport in the WLS fibers, and photon detection by SiPMs.
The SiPM response was modeled by converting detected optical photons to photoelectrons, including the effects of dark counts, cross-talk, afterpulsing, and saturation due to the finite number of SiPM pixels.
The purpose of the simulation in this work is not to describe every optical process in full detail, but to reproduce the observables most relevant to the position reconstruction, namely the channel-to-channel light yield distribution and the total detected light yield.
The simulation parameters relevant to these observables are described below.

The embedded scatterers were modeled as Rayleigh scattering with a scattering length $\lambda_{\mathrm{scat}}$.
This scattering process was introduced as an effective model of the effect of the scatterers on the lateral spread of scintillation light, rather than being derived from the scatterer particle size.
In practice, absorption of scintillation photons by the scatterers would manifest as a shorter effective absorption length in the scintillator.
However, its impact on the detected light yield is difficult to disentangle from a change in the intrinsic scintillation photon yield $Y_\mathrm{scint}$.
We therefore fixed the scintillator absorption length to 380~cm, which is sufficiently long compared to the scattering length of order millimeters (Sec.~\ref{beamtest:simtuning}) to ensure that absorption in the scintillator does not affect the tuning of $\lambda_{\mathrm{scat}}$, and treated $Y_\mathrm{scint}$ as the sole parameter controlling the overall light yield scale.
For the reflective coating on the scintillator surfaces, the effective reflectivity in the assembled detector may differ from the nominal value, which can lead to data--simulation differences in the light yield distribution, particularly near the scintillator edges.
We therefore treated the reflectivity as a tunable parameter.
Accordingly, to reproduce the measured response, three optical parameters were tuned in the simulation: $\lambda_{\mathrm{scat}}$, $Y_\mathrm{scint}$, and the reflectivity of the reflective coating on the scintillator surfaces.
Other optical properties, such as the absorption and emission spectra of the WLS fibers and the refractive indices of the optical materials, were taken from manufacturer datasheets and kept fixed.
The tuning procedure and the simulation configuration for the beam test are described in Sec.~\ref{beamtest:simtuning}.

\begin{figure}[htbp]
  \centering
  \includegraphics[width=0.65\linewidth]{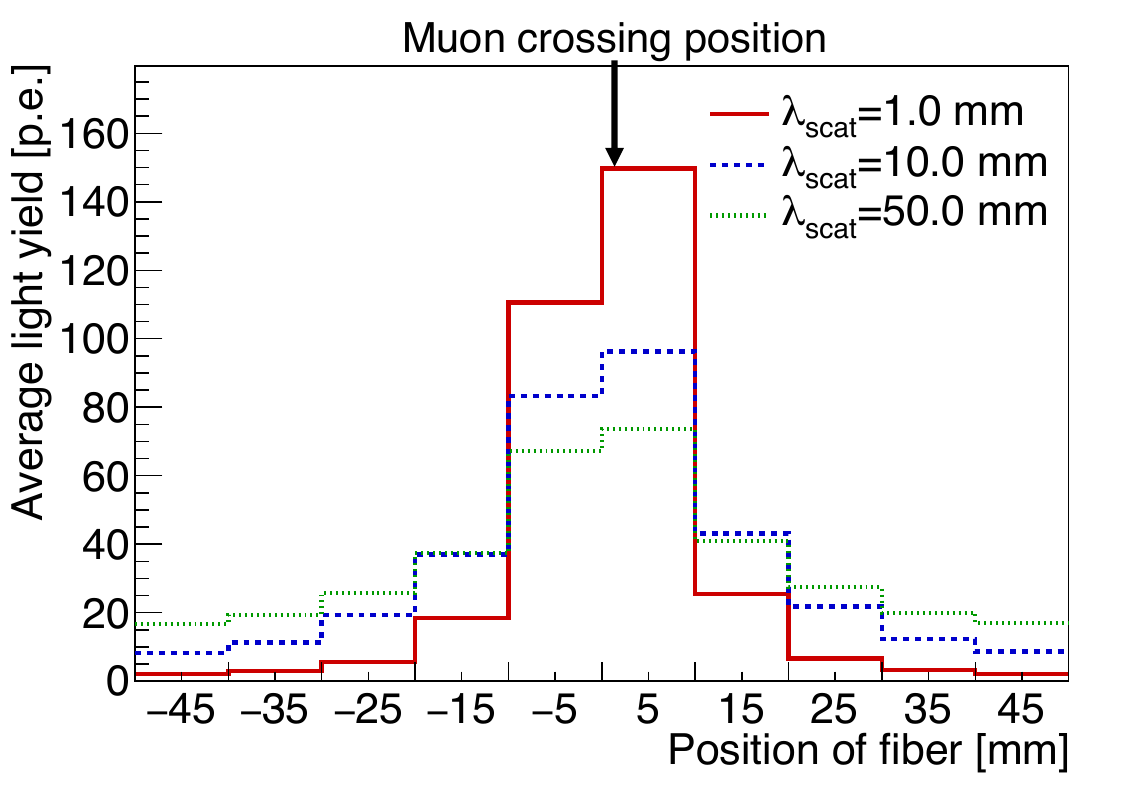}
  \caption{Event-averaged light yield distributions in the $x$ fiber array for normally incident 1~GeV muons injected at $(x,y)=(1~\mathrm{mm},0~\mathrm{mm})$, shown for $\lambda_{\mathrm{scat}}=1.0~\mathrm{mm}$, $10.0~\mathrm{mm}$, and $50.0~\mathrm{mm}$.}
  \label{fig:simulation:lightdistribution}
\end{figure}

To examine the detector concept, we simulated how the light yield distribution changes with the scattering length $\lambda_{\mathrm{scat}}$.
The simulation used a scintillator plate with an area of $100~\mathrm{mm}\times100~\mathrm{mm}$ and a thickness of $10~\mathrm{mm}$.
WLS fibers were arranged at a pitch of $10~\mathrm{mm}$, the scintillation photon yield was fixed at $Y_\mathrm{scint}=7000$~photons/MeV, and the reflectivity of the reflective paint was taken from the datasheet of Eljen EJ-510~\cite{eljen:EJ510}.
Normally incident 1~GeV muons were injected at $(x,y)=(1~\mathrm{mm},0~\mathrm{mm})$, taking the origin of the coordinate system $(x,y)=(0~\mathrm{mm},0~\mathrm{mm})$ to be at the center of the plate.
For each of $\lambda_{\mathrm{scat}}=1.0~\mathrm{mm}$, $10.0~\mathrm{mm}$, and $50.0~\mathrm{mm}$, 10\,000 events were simulated.
Figure~\ref{fig:simulation:lightdistribution} shows the event-averaged light yield distributions in the $x$ fiber array.
The horizontal axis represents the fiber positions, and the vertical axis represents the average light yield in each readout channel.
As expected, the light yield is larger in channels closer to the injection position, and the distribution becomes more localized as $\lambda_{\mathrm{scat}}$ decreases.

\section{Position reconstruction and simulation study}\label{sec:posrec}
\subsection{Position reconstruction method}
To quantitatively extract the particle crossing position,
we developed a reconstruction algorithm that maps the observed light yield distribution to the position.

\begin{figure}[htbp]
  \centering
  \includegraphics[width=0.6\linewidth]{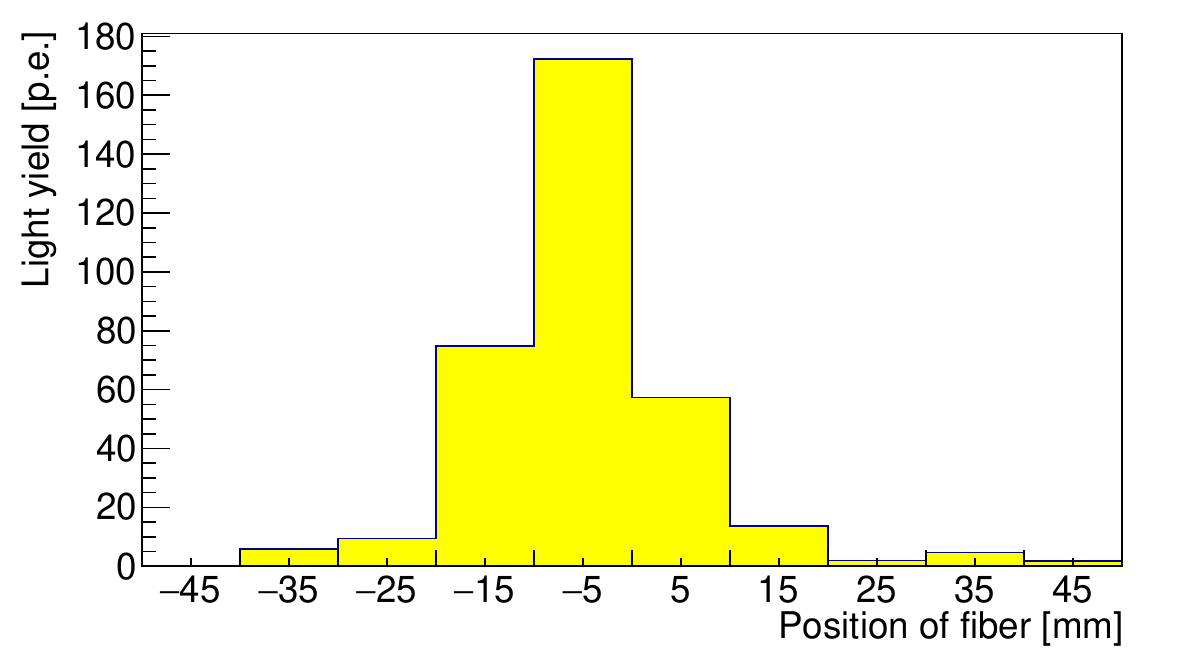}
  \caption{Example of the light yield distribution obtained in the beam test.}
  \label{fig:lightyielddistribution}
\end{figure}

Figure~\ref{fig:lightyielddistribution} shows an example of the light yield distribution measured in the beam test described in Sec.~\ref{sec:beamtest}.
To convert this light yield distribution into a position-sensitive quantity, we use the weighted center of light yield $x_g$, defined as
\begin{align}
\label{eq:xgdef}
x_g=\sum_{i}\frac{N_{i}^{\mathrm{obs}}x_{i}}{N}~\left(N=\sum_{i}N_{i}^{\mathrm{obs}}\right),
\end{align}
where $i$ is the channel index, $x_i$ is the fiber position of the $i$th channel, and $N_{i}^{\mathrm{obs}}$ is the observed light yield in that channel.
For the example in Fig.~\ref{fig:lightyielddistribution}, Eq.~\eqref{eq:xgdef} gives $x_g=-4.8~\mathrm{mm}$.

\begin{figure}[t]
  \centering
  \begin{subfigure}{0.6\textwidth}
    \centering
    \includegraphics[width=\linewidth]{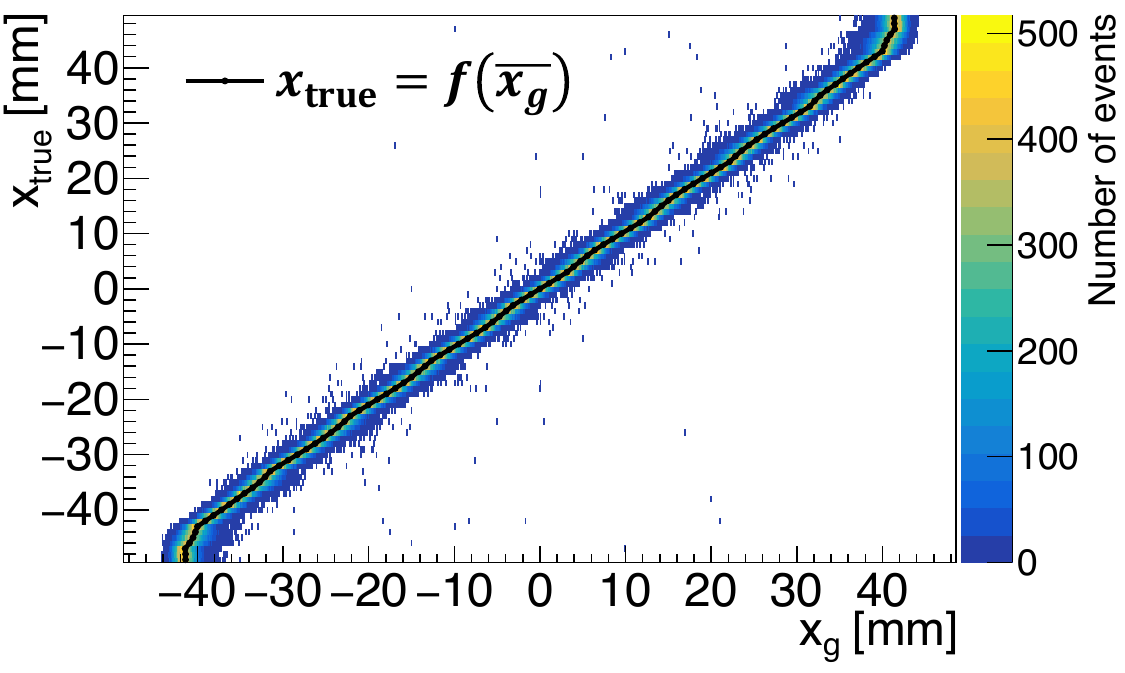}
    \caption{Full range.}
    \label{fig:correctionfunction:full}
  \end{subfigure}
  \hfill
  \begin{subfigure}{0.48\textwidth}
    \centering
    \includegraphics[width=\linewidth]{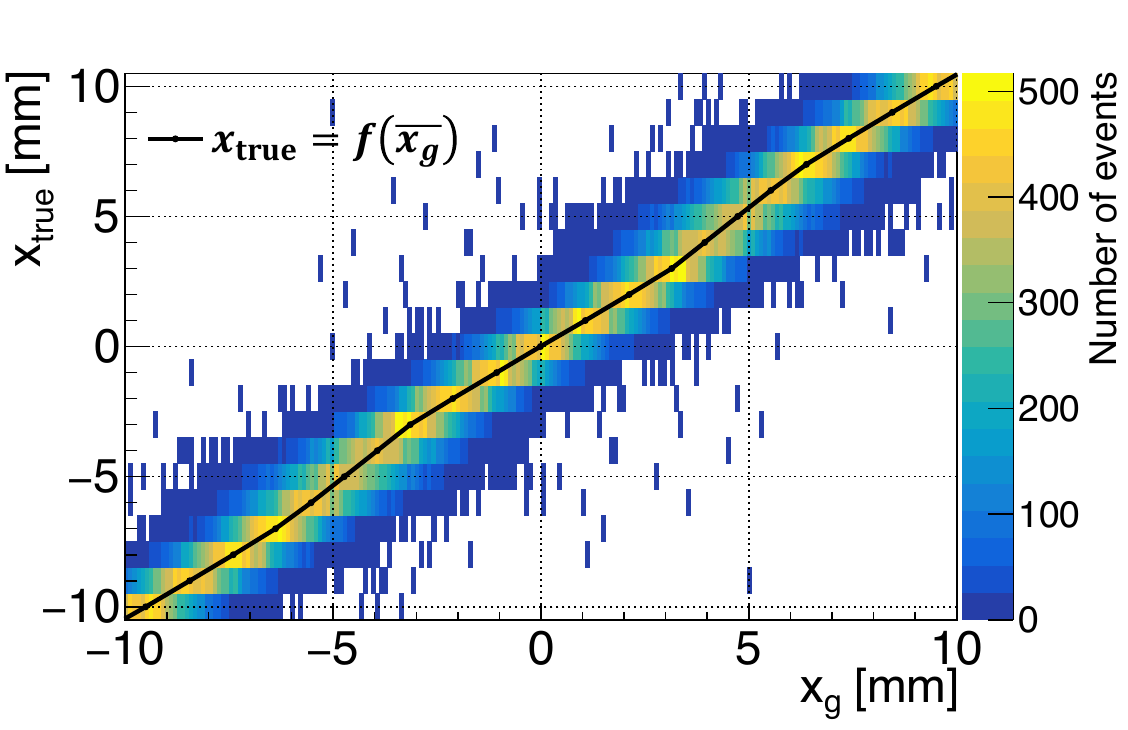}
    \caption{Central region, $-10~\mathrm{mm} \le x_g \le 10~\mathrm{mm}$.}
    \label{fig:correctionfunction:center}
  \end{subfigure}
  \hfill
  \begin{subfigure}{0.48\textwidth}
    \centering
    \includegraphics[width=\linewidth]{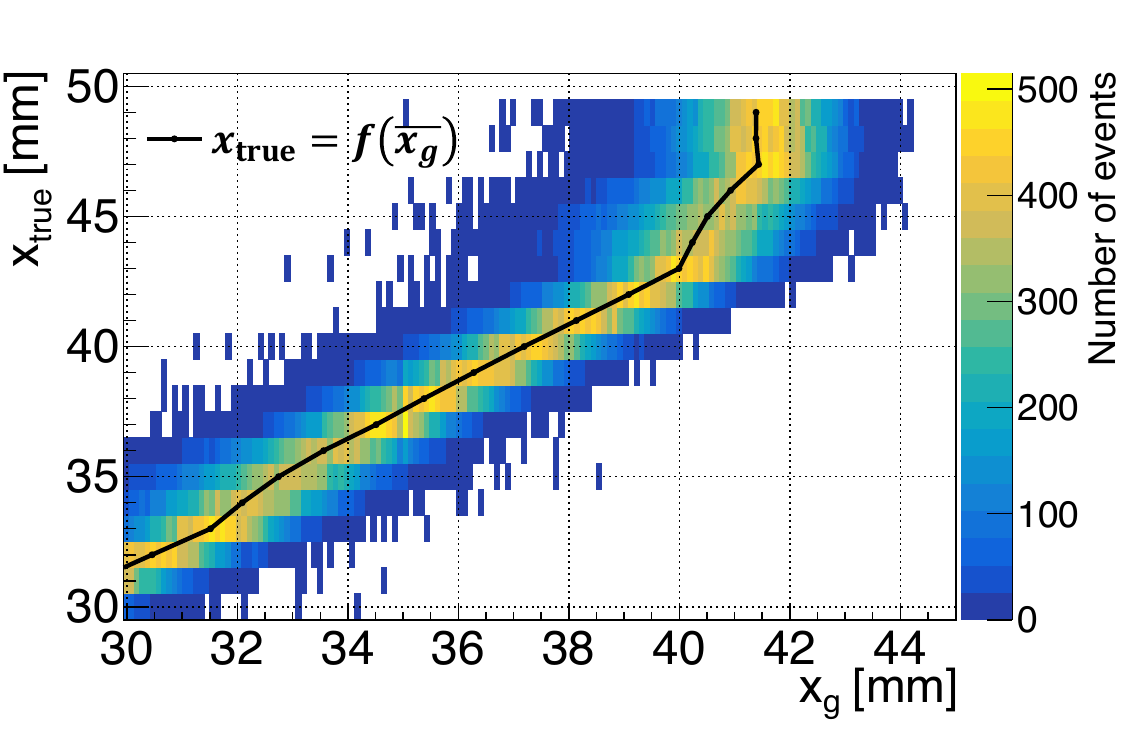}
    \caption{Edge region, $30~\mathrm{mm} \le x_g \le 45~\mathrm{mm}$.}
    \label{fig:correctionfunction:edge}
  \end{subfigure}
  \caption{Mapping function $x_{\mathrm{true}}=f(\overline{x_g})$ used in the beam test, overlaid on a 2D-histogram color map of $x_g$ as a function of $x_{\mathrm{true}}$.}
  \label{fig:correctionfunction}
\end{figure}

While $x_g$ is strongly correlated with the true position, it is not strictly equal to it due to geometric and optical effects, such as the finite fiber diameter, the groove structure, the finite fiber pitch, and photon reflections at the scintillator side surfaces.
We correct for this discrepancy using a mapping function, which relates the average weighted center of light yield $\overline{x_g}$ to the true position $x_{\mathrm{true}}$:
\begin{equation}
x_{\mathrm{true}} = f\!\left(\overline{x_g}\right).
\end{equation}
The mapping function $f$ is obtained from an optical simulation described in Sec.~\ref{sec:simulation}.
It is constructed by injecting normally incident 1~GeV muons at known positions, computing $\overline{x_g}$ for each $x_{\mathrm{true}}$, and defining $f$ by linear interpolation of the resulting relation between $x_{\mathrm{true}}$ and $\overline{x_g}$.
The reconstructed position is then obtained for each event as $x_{\mathrm{rec}}=f(x_g)$.
Figure~\ref{fig:correctionfunction}(\subref{fig:correctionfunction:full}) shows the resulting mapping function, overlaid on a 2D-histogram color map of $x_g$ as a function of $x_{\mathrm{true}}$.
Figures~\ref{fig:correctionfunction}(\subref{fig:correctionfunction:center}) and \ref{fig:correctionfunction}(\subref{fig:correctionfunction:edge}) show zoomed-in views of the central and edge regions, respectively.
Using this mapping function, the example event in Fig.~\ref{fig:lightyielddistribution} yields $x_{\mathrm{rec}}=-5.1~\mathrm{mm}$ for $x_g=-4.8~\mathrm{mm}$.

It should be noted that the kinks at $\overline{x_g}=\pm40~\mathrm{mm}$ in Fig.~\ref{fig:correctionfunction}(\subref{fig:correctionfunction:full}) arise from reflections of scintillation photons at the side surfaces of the $100~\mathrm{mm}\times100~\mathrm{mm}$ scintillator.
As seen more clearly in the edge region shown in Fig.~\ref{fig:correctionfunction}(\subref{fig:correctionfunction:edge}), the slope of $x_{\mathrm{true}}=f(\overline{x_g})$ becomes steeper near the scintillator edges.
Therefore, a given fluctuation in $x_g$ translates into a larger variation in $x_{\mathrm{rec}}$, resulting in degraded position resolution near the scintillator edges.

\subsection{Simulation-based evaluation of the position resolution}\label{sec:posrec:simstudy}

\begin{figure}[htbp]
  \centering
  \begin{subfigure}{0.48\textwidth}
    \centering
    \includegraphics[width=\linewidth]{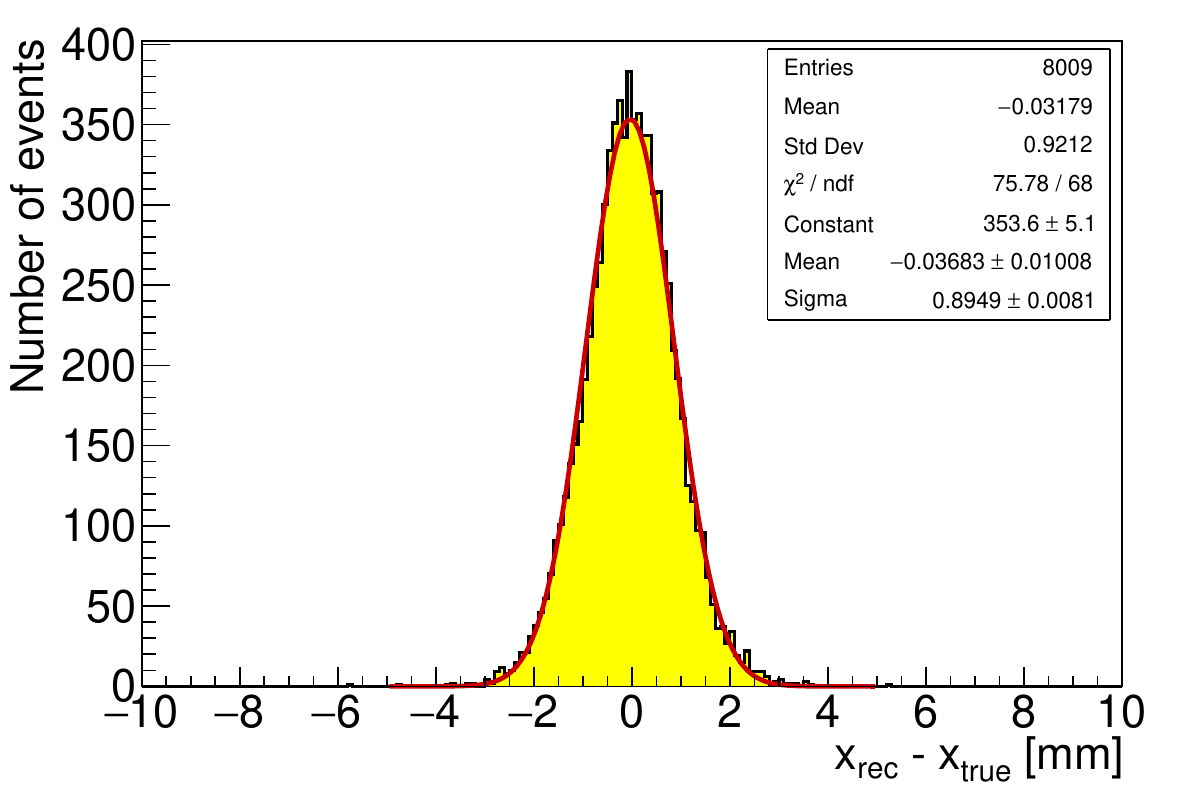}
    \caption{$x_{\mathrm{rec}}-x_{\mathrm{true}}$.}
    \label{fig:posrec:resolutionfit:after}
  \end{subfigure}
  \hfill
  \begin{subfigure}{0.48\textwidth}
    \centering
    \includegraphics[width=\linewidth]{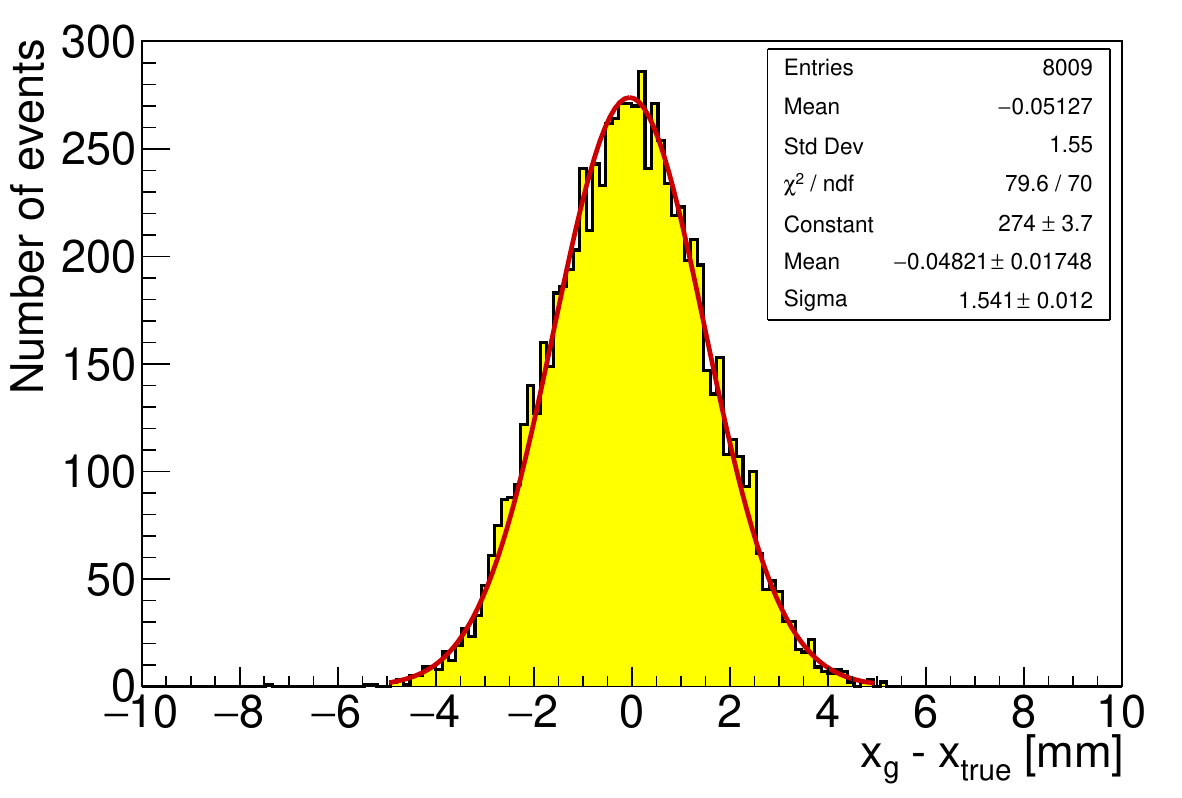}
    \caption{$x_g-x_{\mathrm{true}}$.}
    \label{fig:posrec:resolutionfit:before}
  \end{subfigure}
  \caption{Residual distributions for $\lambda_{\mathrm{scat}}=1.0~\mathrm{mm}$ and $Y_\mathrm{scint}=7000$~photons/MeV in the fiducial region $-40~\mathrm{mm}<x_{\mathrm{true}}<40~\mathrm{mm}$.}
  \label{fig:posrec:resolutionfit}
\end{figure}

\begin{figure}[hbtp]
  \centering
  \includegraphics[width=0.6\linewidth]{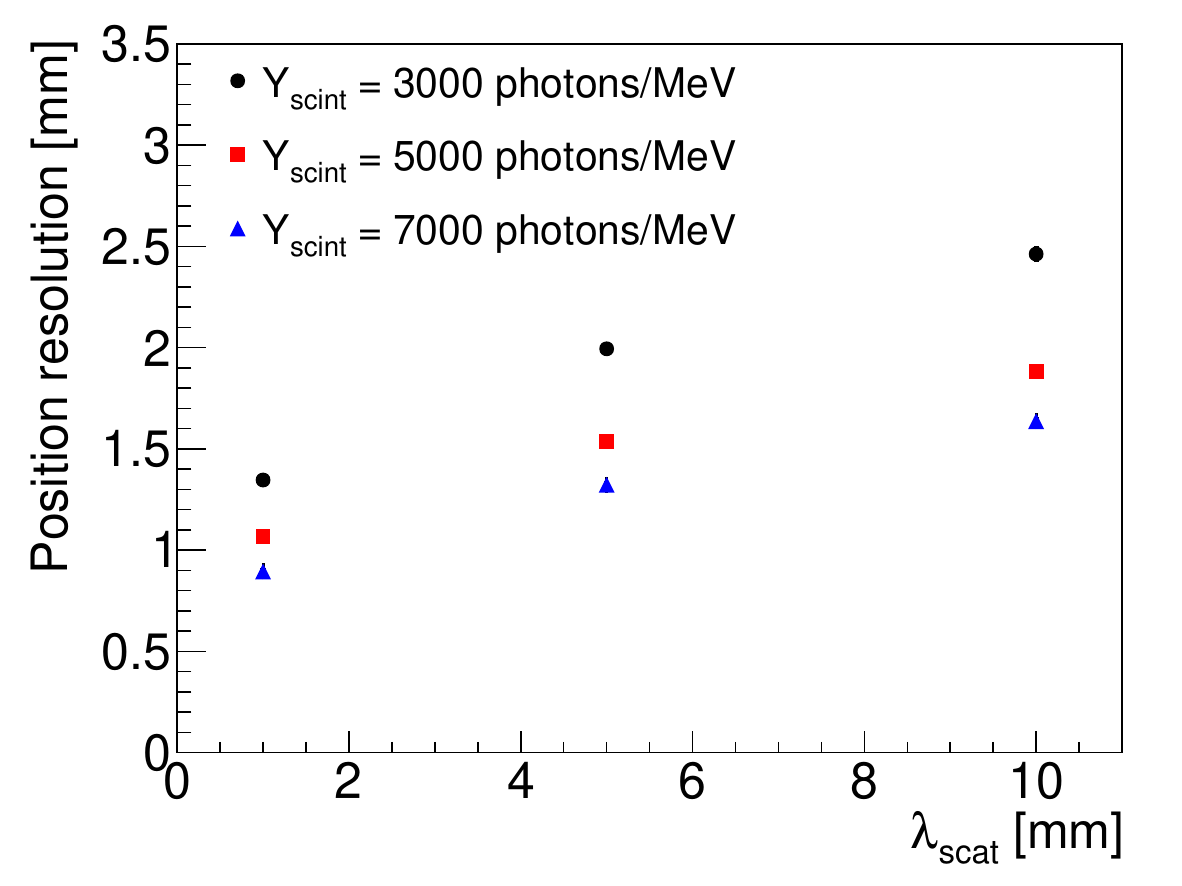}
  \caption{Simulated position resolution as a function of $\lambda_{\mathrm{scat}}$ for $Y_\mathrm{scint}=3000$, $5000$, and $7000$~photons/MeV.}
  \label{fig:posrec:resolution}
\end{figure}

Using the optical simulation, we evaluated the expected position resolution for $\lambda_{\mathrm{scat}}=1.0$, $5.0$, and $10.0~\mathrm{mm}$ and $Y_\mathrm{scint}=3000$, $5000$, and $7000$~photons/MeV.
The same detector configuration as that used for the light yield distributions in Fig.~\ref{fig:simulation:lightdistribution} was adopted.
Normally incident 1~GeV muons were injected at random positions over the $100~\mathrm{mm}\times100~\mathrm{mm}$ scintillator, and the reconstructed positions were obtained using the procedure described above.
To avoid the degradation in position reconstruction near the scintillator edges, the position resolution was evaluated in the fiducial region $-40~\mathrm{mm}<x_{\mathrm{true}}<40~\mathrm{mm}$, excluding the outermost 10~mm from the edges.
The position reconstruction was found to work well in all simulated cases.
The position resolution was evaluated from the width of a Gaussian fit to the residual distribution $x_{\mathrm{rec}}-x_{\mathrm{true}}$.
Figure~\ref{fig:posrec:resolutionfit} shows the residual distributions and the Gaussian fits for $\lambda_{\mathrm{scat}}=1.0~\mathrm{mm}$ and $Y_\mathrm{scint}=7000$~photons/MeV.
The distribution of $x_g-x_{\mathrm{true}}$, obtained before applying the mapping function, is also shown for comparison. Applying the mapping function reduced the fitted residual width from 1.54~mm to 0.89~mm, corresponding to a factor of 1.73 reduction.

Figure~\ref{fig:posrec:resolution} summarizes the simulated position resolution for the values of $\lambda_{\mathrm{scat}}$ and $Y_\mathrm{scint}$ considered in this study.
For all three values of $Y_\mathrm{scint}$, the position resolution improves as $\lambda_{\mathrm{scat}}$ decreases.
For a fixed $\lambda_{\mathrm{scat}}$, the position resolution also improves as $Y_\mathrm{scint}$ increases.
These results show that both sufficiently localized scintillation light and sufficient detected light yield are important for achieving a position resolution well below the readout pitch.

\subsection{Remarks on multiple-particle hits}

\begin{figure}[htbp]
  \centering
  \begin{subfigure}{0.48\textwidth}
    \centering
    \includegraphics[width=\linewidth]{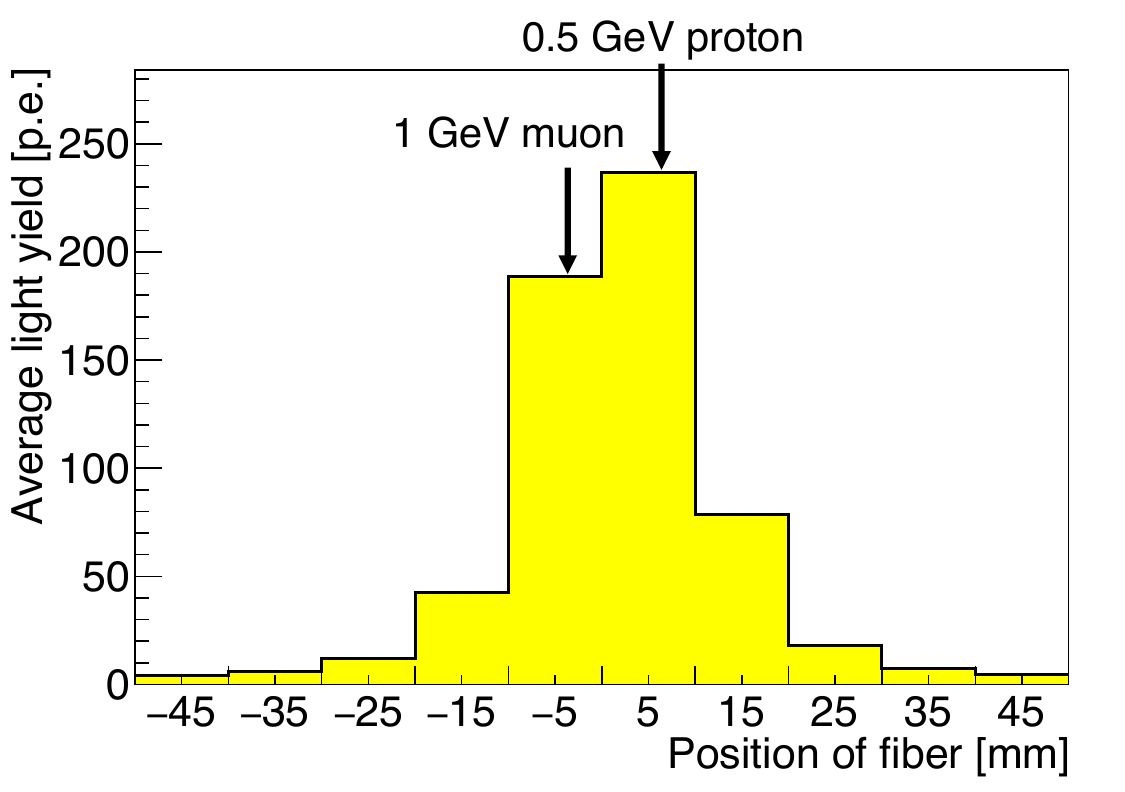}
    \caption{Proton at $(x,y)=(6~\mathrm{mm},0~\mathrm{mm})$.}
  \end{subfigure}
  \hfill
  \begin{subfigure}{0.48\textwidth}
    \centering
    \includegraphics[width=\linewidth]{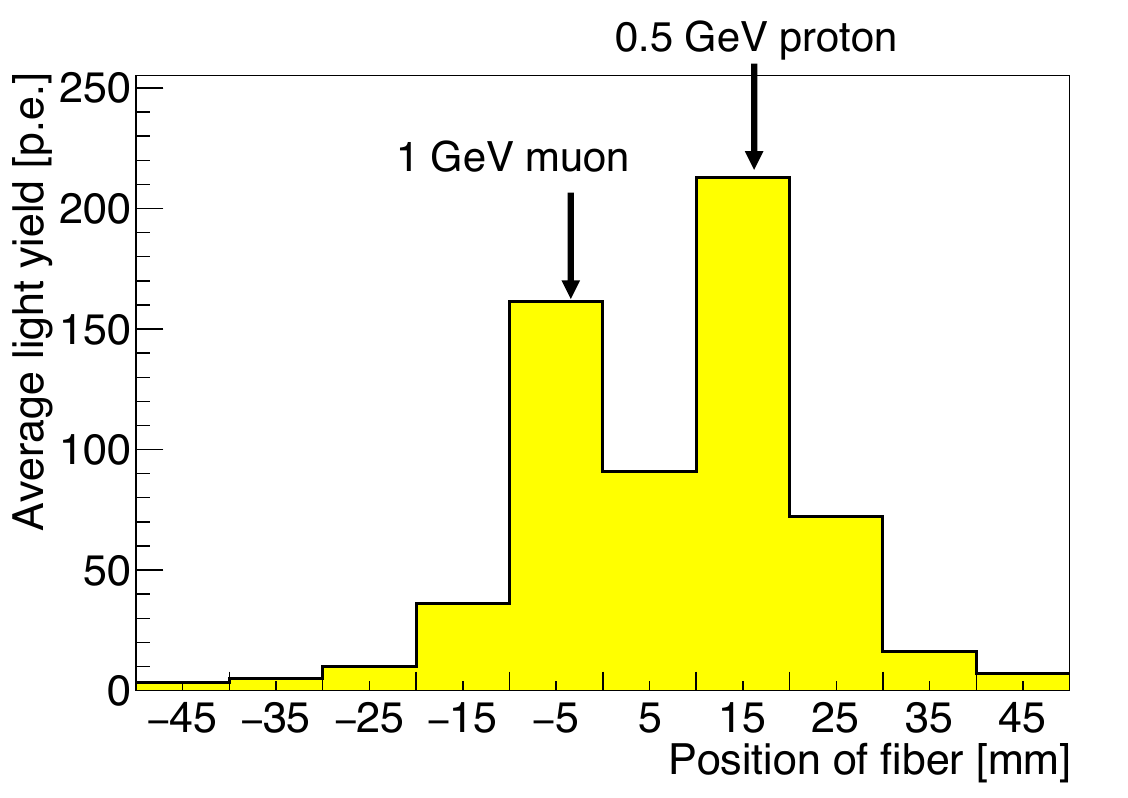}
    \caption{Proton at $(x,y)=(16~\mathrm{mm},0~\mathrm{mm})$.}
  \end{subfigure}
  \caption{Event-averaged light yield distributions in the $x$ fiber array for simulated multiple-particle hits. A normally incident 1~GeV muon was injected at $(x,y)=(-4~\mathrm{mm},0~\mathrm{mm})$, together with a normally incident 0.5~GeV proton injected at the position indicated in each panel.}
  \label{fig:posrec:multihit}
\end{figure}

Although this paper focuses on the single-particle case, we briefly comment on the impact of multiple-particle hits on the position reconstruction.
Figure~\ref{fig:posrec:multihit} shows the event-averaged light yield distributions in the $x$ fiber array for simulated multiple-particle hits, in which a normally incident 1~GeV muon was injected at $(x,y)=(-4~\mathrm{mm},0~\mathrm{mm})$ together with a normally incident 0.5~GeV proton injected at $(x,y)=(6~\mathrm{mm},0~\mathrm{mm})$ or $(16~\mathrm{mm},0~\mathrm{mm})$.
For each configuration, 10\,000 events were simulated.

For the 10-mm separation, the distribution forms a single broad peak, whereas for the 20-mm separation, two distinct peaks are visible.
This suggests that dedicated reconstruction algorithms would be required for multiple-particle hits, and that the achievable position resolution would be worse than in the single-particle case.
The degradation is expected to be particularly large when the individual contributions are not clearly separated in the light yield distribution, as in the 10-mm case.

\section{Performance evaluation of FROST with a positron beam test using prototype detectors}
\label{sec:beamtest}

\subsection{Beam test overview and objectives}
\label{beamtest:overview}

A positron beam test was performed at RARiS~\cite{raris:Hama:2020}, Tohoku University, in July 2024.
As shown in Sec.~\ref{sec:posrec:simstudy}, the optical simulation indicates that position reconstruction based on the light yield distribution works under idealized conditions.
In practice, however, various effects such as detector non-uniformity, optical imperfections, and electronic noise can degrade the reconstruction performance.
The beam test was therefore carried out with FROST prototypes exposed to a 730~MeV positron beam in order to validate this reconstruction principle experimentally and to evaluate the detector performance of FROST under realistic conditions.
The beam test was carried out with the objectives listed below.
\begin{enumerate}
\item \textbf{Evaluation of the detection efficiency.}\\
As explained in Sec.~\ref{sec:intro}, FROST is expected to provide a near-100\% detection efficiency in principle.
We evaluated the detection efficiency of the FROST prototypes.

\item \textbf{Evaluation of the position resolution and its angular dependence.}\\
We evaluated the position resolution of FROST prototypes for normally incident positrons and for inclined incidence at several angles up to $45^\circ$ to assess its angular dependence.

\item \textbf{Optimization of the scatterer concentration.}\\
Localization of the scintillation light is essential for position reconstruction in FROST and is intended to be achieved by embedding scatterers in the scintillator.
However, an excessive scatterer concentration can reduce the detected light yield due to absorption of scintillation photons by the scatterers, which may degrade the position resolution.
The beam test therefore compared prototypes with different scatterer concentrations to determine the optimal scatterer concentration that optimizes the position resolution.

\item \textbf{Investigation of the effect of scintillator bonding.}\\
Although the FROST concept is based on a monolithic scintillator plate, fabricating a physically monolithic plastic scintillator with an area of the order of \(\mathrm{m}^2\) or larger is practically challenging.
A realistic approach to large-area instrumentation is therefore to assemble the plate by bonding multiple scintillator tiles using optical cement.
Such bonding inevitably introduces internal boundaries, which may affect both the position resolution and detection efficiency.
In the beam test, we investigated these effects of the bonded interfaces.
\end{enumerate}

\subsection{Beamline at RARiS}

An electron beam accelerated to approximately 90~MeV by a linear accelerator is injected into a synchrotron called Booster STorage (BST), where it is further accelerated up to a maximum energy of 1.3~GeV.
A carbon fiber inserted into the orbit of the circulating electron beam produces $\gamma$ rays via bremsstrahlung~\cite{raris:Ishikawa:2010}.
These $\gamma$ rays are then directed onto a tungsten target, where electron--positron pairs are generated through pair production.
The positrons are selectively bent by a dipole magnet and transported to the test beam line at the GeV-$\gamma$ experimental hall~\cite{raris:Ishikawa:2012}, where the detector setup described in Sec.~\ref{beamtest:detsetup} was installed.
A 730~MeV positron beam was selected for this beam test.
The transverse beam profiles were evaluated using the hodoscopes described later, yielding beam widths of $18~\mathrm{mm}$ and $11~\mathrm{mm}$ in the horizontal and vertical directions, respectively.
With the trigger configuration described in Sec.~\ref{beamtest:detsetup}, the average event rate during data taking was about 500~Hz.

\subsection{Detector setup}
\label{beamtest:detsetup}
\begin{figure}[htbp]
  \centering
  \begin{subfigure}{0.4\textwidth}
    \centering
    \includegraphics[width=\linewidth]{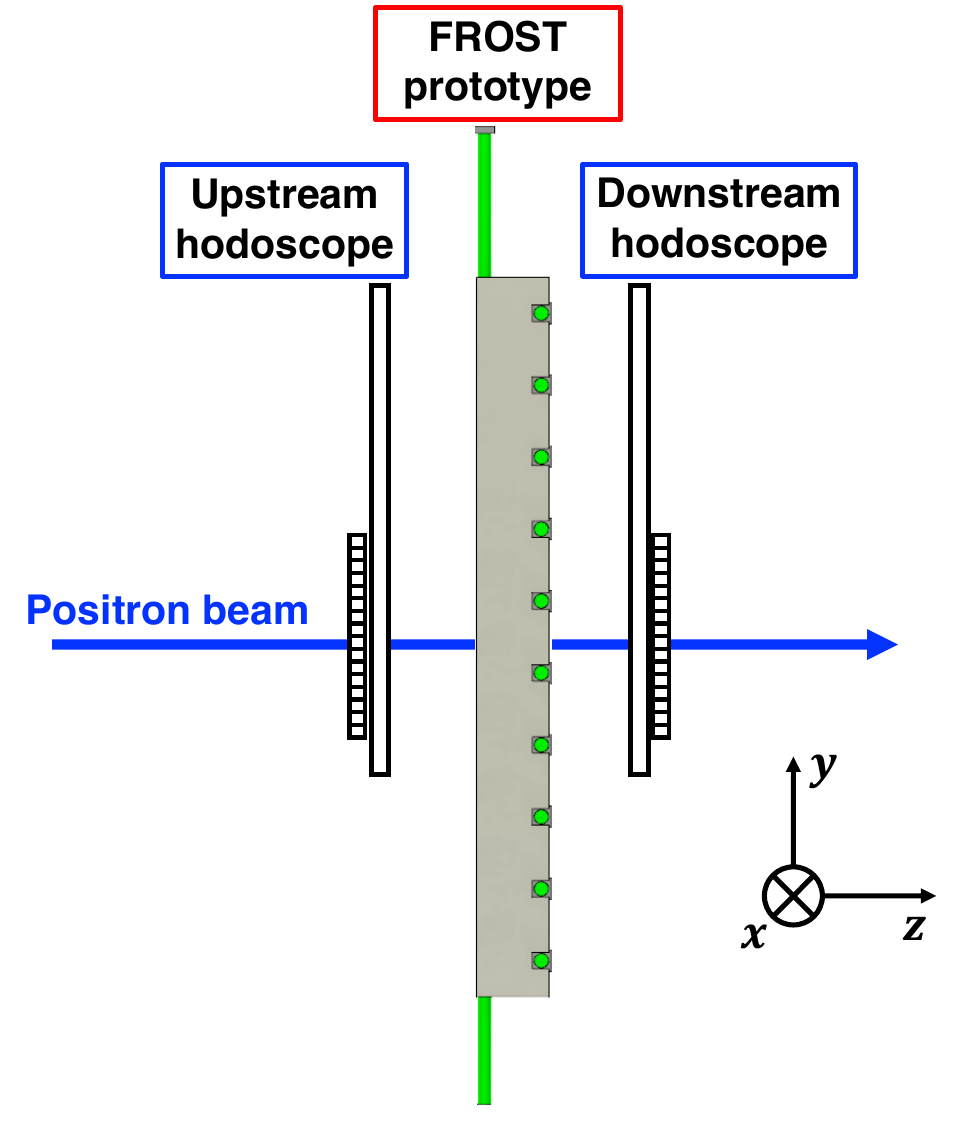}
    \caption{Schematic view of the detector setup.}
    \label{fig:beamtest:setupgainenzu}
  \end{subfigure}
  \hfill
  \begin{subfigure}{0.58\textwidth}
    \centering
    \includegraphics[width=\linewidth]{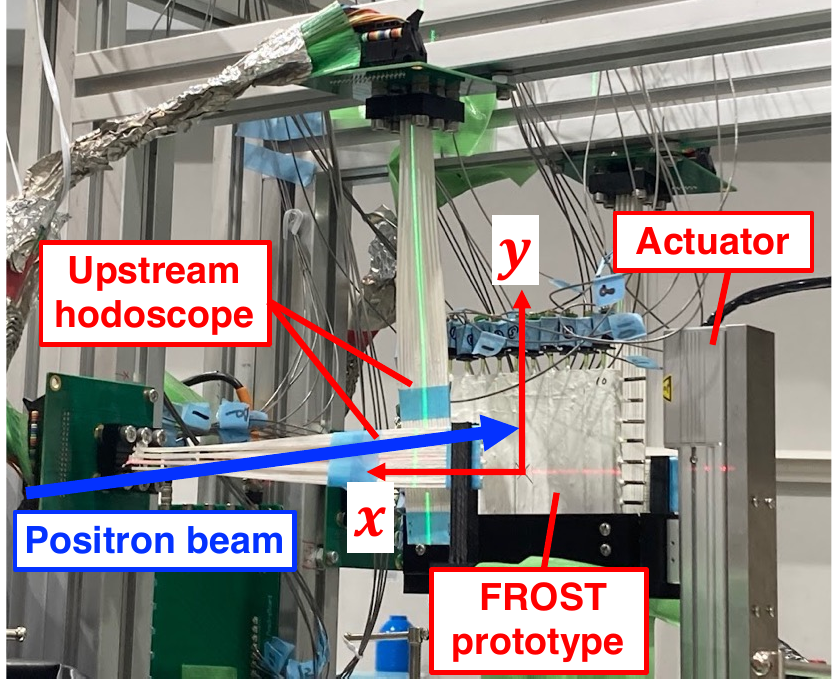}
    \caption{Photograph of the detector setup.}
    \label{fig:beamtest:setuppicture}
  \end{subfigure}
  \caption{The setup of the beam test at RARiS.}
  \label{fig:beamtest:setup}
\end{figure}

Figure~\ref{fig:beamtest:setup} shows the detector setup used in the beam test.
A prototype FROST detector was placed between an upstream and a downstream hodoscope.
The hodoscopes provided a reference measurement of the positron crossing position, which was used to determine the beam position on the FROST prototype.
The FROST prototype was mounted on an actuator, which allowed us to move it in the horizontal and vertical directions and change the beam irradiation position on the FROST prototype.
The right-handed coordinate system was defined with the origin at the center of the FROST prototype; the $z$ axis was along the beam direction, and the $x$ and $y$ axes were the horizontal and vertical transverse directions, respectively.
In the following, the FROST prototypes and the hodoscopes are described in detail.

\subsubsection{FROST prototypes}
\label{beamtest:prototype}

\begin{figure}[htbp]
  \centering
  \includegraphics[width=\linewidth]{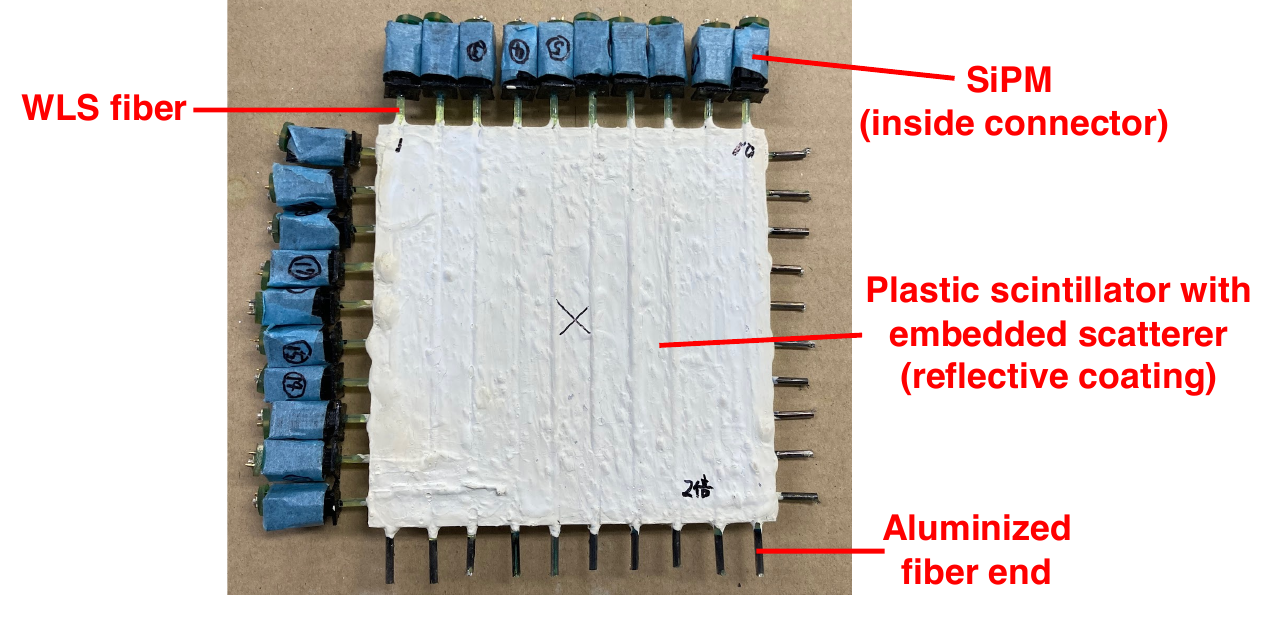}
  \caption{Photograph of a FROST prototype detector.}
  \label{fig:beamtest:prototype_picture}
\end{figure}

Each FROST prototype consisted of a plastic scintillator plate of $100~\mathrm{mm}\times100~\mathrm{mm}$ area and 10~mm thickness, with embedded scatterers.
The scintillator plates with embedded scatterers were custom-fabricated by Kuraray.
The scintillator was based on the same base material as Kuraray SCSF-78, while its detailed formulation differs from that of standard SCSF-78. Details of the fluorescent dopants, including their species and concentrations, as well as the scatterer material, particle size, and absolute scatterer concentration, are confidential to the manufacturer and cannot be disclosed.
Grooves with a width and depth of 2.5~mm were machined on both scintillator surfaces in orthogonal directions at a pitch of 10~mm, and WLS fibers with a diameter of 2.0~mm (Kuraray Y-11(200)M, 2.0mmD, BSJ) were embedded in the grooves and optically coupled to the scintillator using optical cement (Eljen EJ-500).
The fiber positions were $x,y=-45,-35,\cdots,45~\mathrm{mm}$.
Each WLS fiber was coupled to a SiPM (Hamamatsu S13360-3075CS), operated at approximately 4~V above the breakdown voltage, and the SiPM signal waveforms were recorded with a digitizer (CAEN DT5740).
The scintillator surfaces were coated with a $\mathrm{TiO_2}$-based reflective paint (Eljen EJ-510) to enhance light collection.
To further increase the collected light yield, the fiber end opposite to the SiPM was aluminized by vapor deposition to form a reflective mirror.
These features of the prototypes were adopted to increase the collected light yield, which is essential for achieving a position resolution below the 10-mm readout pitch.
The photograph of a FROST prototype is shown in Fig.~\ref{fig:beamtest:prototype_picture}.

\begin{figure}[htbp]
  \centering
  \includegraphics[width=0.8\linewidth]{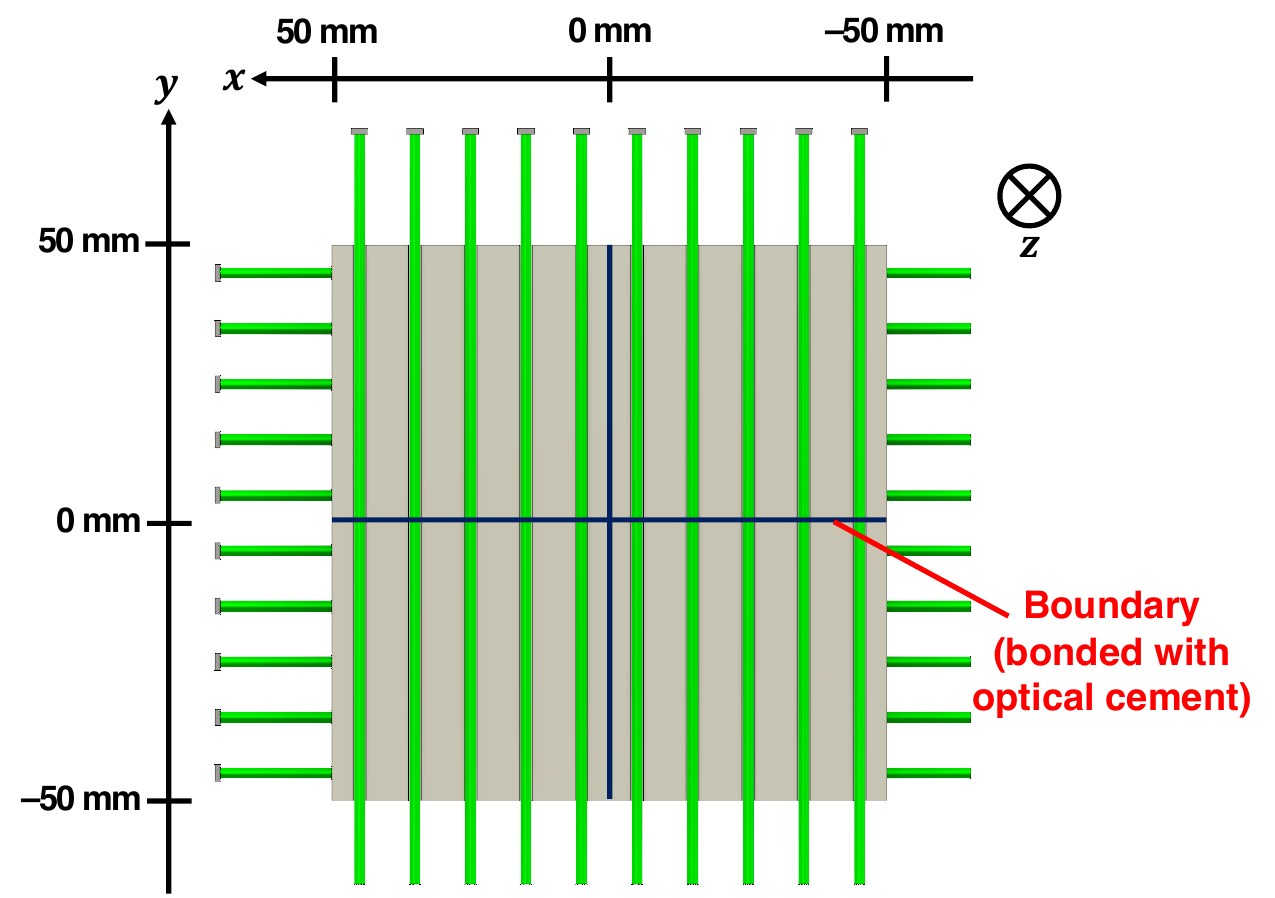}
  \caption{Schematic view of Prototype~T2 assembled from four scintillator tiles bonded with optical cement.}
  \label{fig:beamtest:prototype_combination}
\end{figure}

Four prototypes were prepared for the beam test.
Three detectors were fabricated from monolithic scintillator plates with different scatterer concentrations (Prototypes~M1--M3), and one detector was assembled by bonding four $50~\mathrm{mm}\times50~\mathrm{mm}\times10~\mathrm{mm}$ tiles with optical cement (Prototype~T2; see the schematic view in Fig.~\ref{fig:beamtest:prototype_combination}).
Table~\ref{tab:beamtest:prototypenoudo} summarizes the scintillator configuration of each prototype and the scatterer concentration relative to that of Prototype~M1.

\begin{table}[H]
  \centering
  \caption{Scintillator configuration and relative scatterer concentration of the prototype detectors used in the beam test.}
  \label{tab:beamtest:prototypenoudo}
  \begin{tabular}{c|c c} \hline\hline
  Prototype & Scintillator configuration & Relative scatterer concentration \\\hline
  M1 & Monolithic scintillator plate & 1 \\
  M2 & Monolithic scintillator plate & 2 \\
  M3 & Monolithic scintillator plate & 3 \\
  T2 & \makecell{Four scintillator tiles\\ bonded with optical cement} & 2 \\\hline\hline
 \end{tabular}
\end{table}

\subsubsection{Hodoscopes}
\label{beamtest:hodoscope}

Each hodoscope comprised two orthogonal layers ($x$ and $y$), each consisting of a linear array of 16 plastic scintillating fibers (Kuraray SCSF-78SJ).
The fibers had a square core cross section of $1.5\times1.5~\mathrm{mm}^2$ and were manually coated with a $\mathrm{TiO_2}$-based reflective paint (Eljen EJ-510) with a thickness of approximately $0.1~\mathrm{mm}$, resulting in an effective fiber size of $1.7\times1.7~\mathrm{mm}^2$.
The intersection of an $x$-layer fiber and a $y$-layer fiber defined a $16\times16$ array of virtual cells, providing a position granularity of $1.7~\mathrm{mm}\times1.7~\mathrm{mm}$.
The scintillation light from the scintillating fibers was detected using four $4\times4$ SiPM arrays (Hamamatsu S13361-3050AE-04).
Signals from the SiPMs were read out with a NIM EASIROC module~\cite{EASIROC}.
Data from all channels were recorded when two or more channels in the upstream hodoscope had signals above 3.5~photoelectrons (p.e.), with the EASIROC readout synchronized to the digitizer used for the FROST prototype.

\subsection{Measurement items}
\label{beamtest:measitems}

To address the objectives described in Sec.~\ref{beamtest:overview}, two types of measurements were performed.
First, the positron beam was incident normal to the detector surface, and position scans were carried out for each of the four FROST prototypes (Prototypes~M1--M3, T2) by translating the prototype in the plane transverse to the beam.
The scanned region covered approximately \(-50~\mathrm{mm}<x<20~\mathrm{mm}\) and \(-10~\mathrm{mm}<y<50~\mathrm{mm}\).

Second, the angular dependence of the position resolution was studied using Prototype~M3 by varying the incidence angle $\theta_x$ in the $x$--$z$ plane, where $\theta_x$ was defined as the angle between the beam direction and the $z$ axis in the $x$--$z$ plane.
The angle was changed by rotating the FROST prototype while keeping the hodoscopes fixed, and data were taken at \(\theta_x=15^\circ\), \(30^\circ\), and \(45^\circ\).
The scanned regions were \(-50~\mathrm{mm}<x<20~\mathrm{mm}\) and \(-10~\mathrm{mm}<y<15~\mathrm{mm}\) for \(\theta_x=15^\circ\) and \(30^\circ\), and \(-50~\mathrm{mm}<x<30~\mathrm{mm}\) and \(-10~\mathrm{mm}<y<50~\mathrm{mm}\) for \(\theta_x=45^\circ\).

In both measurement modes, the beam irradiation position was changed by translating only the prototype in the transverse plane, while the hodoscopes were kept fixed. The translation step sizes were chosen as integer multiples of the hodoscope cell pitch (1.7~mm).
The scanned region was divided into position cells defined by the hodoscope crossing position, which correspond to $1.7~\mathrm{mm}/\cos\theta_x$ in $x$ and $1.7~\mathrm{mm}$ in $y$, and typically 500--2500 events were recorded per position cell.

\subsection{Analysis}

\subsubsection{SiPM calibration and correction for fiber/SiPM variations}
\label{beamtest:calib}

For the FROST prototype data, the integrated waveform charge in analog-to-digital converter (ADC) counts was linearly converted to the number of photoelectrons using the pedestal and the 1-p.e.-equivalent ADC integral obtained from SiPM dark-noise data, and the resulting value was used as the light yield observable.
To reduce channel-to-channel variations arising from individual WLS fibers and SiPMs, we measured the cosmic-ray light yield for each channel before the fibers were bonded to the FROST scintillator, using a dedicated plastic scintillator bar of $50~\mathrm{mm}\times100~\mathrm{mm}\times10~\mathrm{mm}^3$ with holes of approximately 2.2~mm diameter into which the WLS fibers were inserted.
Each fiber was read out with the same SiPM as used for the FROST prototype.
In the analysis, the light yield in each channel was normalized such that this pre-measured cosmic-ray light yield was equalized to the overall mean value of 83.1~p.e.
Before this normalization, the relative channel-to-channel variation was 12.3\%.

\subsubsection{Single-particle event selection and determination of the reference position using the hodoscopes}

\begin{figure}[htbp]
  \centering
  \includegraphics[width=1.0\linewidth]{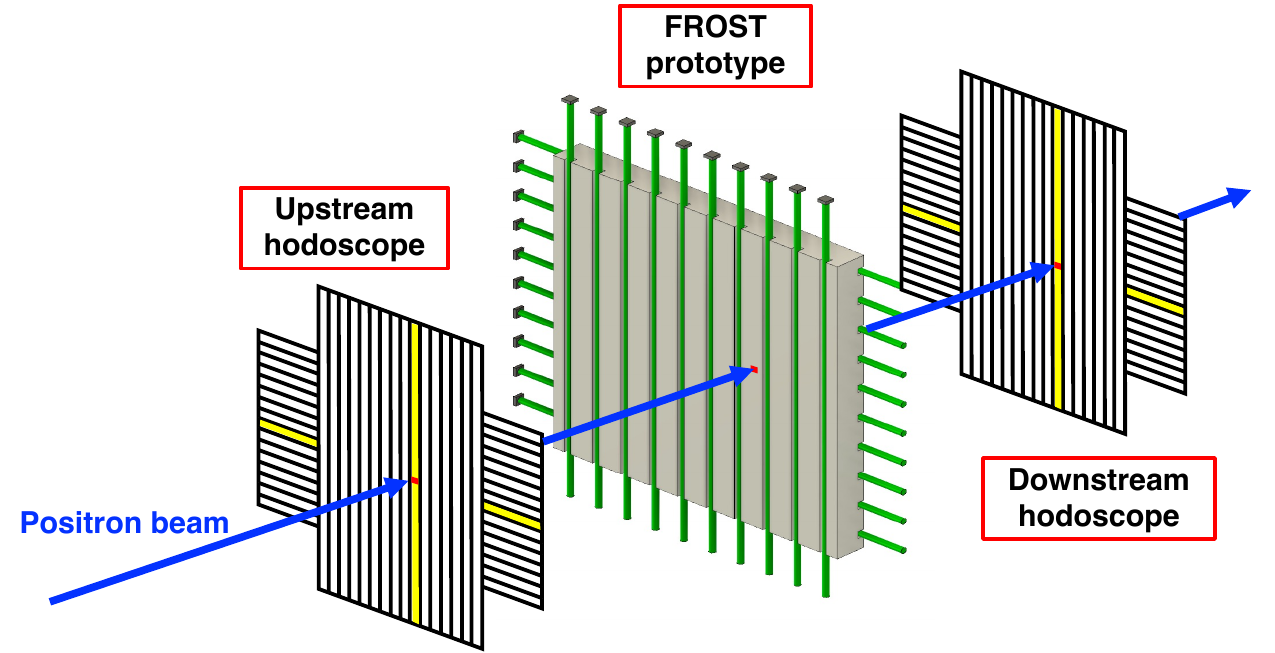}
  \caption{Single-particle event selection with the hodoscopes and the determination of the reference beam position on the FROST prototype.}
  \label{fig:beamtest:eventselection}
\end{figure}

As illustrated in Fig.~\ref{fig:beamtest:eventselection}, the hodoscopes were used to select single-particle events and to determine the reference crossing position on the FROST prototype.
To reject events with multiple particles and to suppress noise hits, we required that exactly one scintillating fiber exceeded a threshold of 3.5~p.e.\ in each of the $x$ and $y$ layers of the upstream hodoscope,
and that at least one fiber exceeded the same threshold in the downstream hodoscope.
For almost all selected events, the downstream hit position was located in the same cell as the upstream hit position or in a neighboring cell.
This confirmed that the effect of scattering in the hodoscopes and in the FROST prototype was sufficiently small.
We therefore took the center of the hit cell in the upstream hodoscope as the reference crossing position on the FROST prototype.

\subsubsection{Simulation setup for the beam test and parameter tuning}
\label{beamtest:simtuning}

The Geant4-based optical simulation framework described in Sec.~\ref{sec:simulation} was configured to model the response of the FROST prototypes in the beam test.
The simulation implemented the full prototype geometry described in Sec.~\ref{beamtest:prototype}.
For Prototype~T2, the four scintillator tiles and the optical-cement boundaries were included in the geometry, and the optical-cement thickness at the tile boundaries was set to 0.05~mm.
The dark-count rate and the probabilities of cross-talk and afterpulsing were determined from SiPM dark-noise measurements and applied channel-by-channel in the simulation.

The three tuned optical parameters were the scattering length $\lambda_{\mathrm{scat}}$, the scintillation photon yield $Y_\mathrm{scint}$, and the reflectivity of the reflective coating (Sec.~\ref{sec:simulation}).
$\lambda_{\mathrm{scat}}$ and $Y_\mathrm{scint}$ were tuned using the normal-incidence data in the central region ($-10~\mathrm{mm}<x_{\mathrm{true}},y_{\mathrm{true}}<10~\mathrm{mm}$).
To reduce biases arising from non-uniform spatial event statistics, the event counts were equalized across the selected position cells.
The reflectivity of the reflective coating was then tuned using the position dependence of the weighted center of light yield over a wider region, as described later.
Details of the tuning procedure are given below for each parameter.

\begin{figure}[htbp]
  \centering
  \begin{subfigure}{0.6\textwidth}
    \centering
    \includegraphics[width=\linewidth]{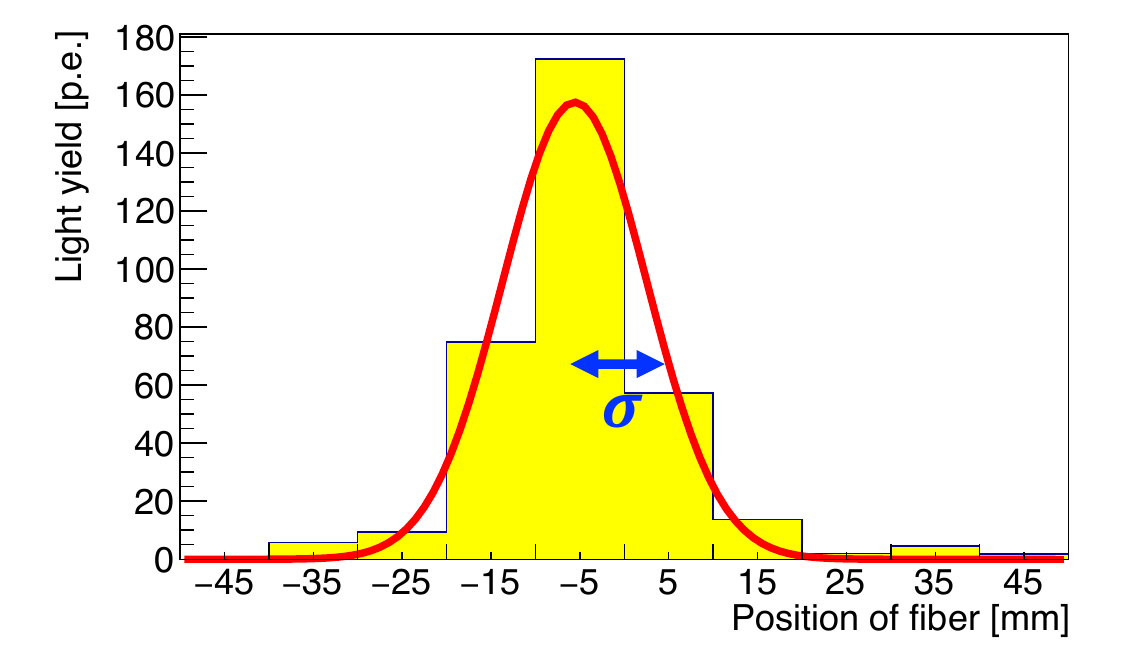}
    \caption{Example of a Gaussian fit to the light yield distribution (Prototype~M3).}
    \label{fig:beamtest:tuning_gauss}
  \end{subfigure}
  \hfill
  \begin{subfigure}{0.95\textwidth}
    \centering
    \includegraphics[width=\linewidth]{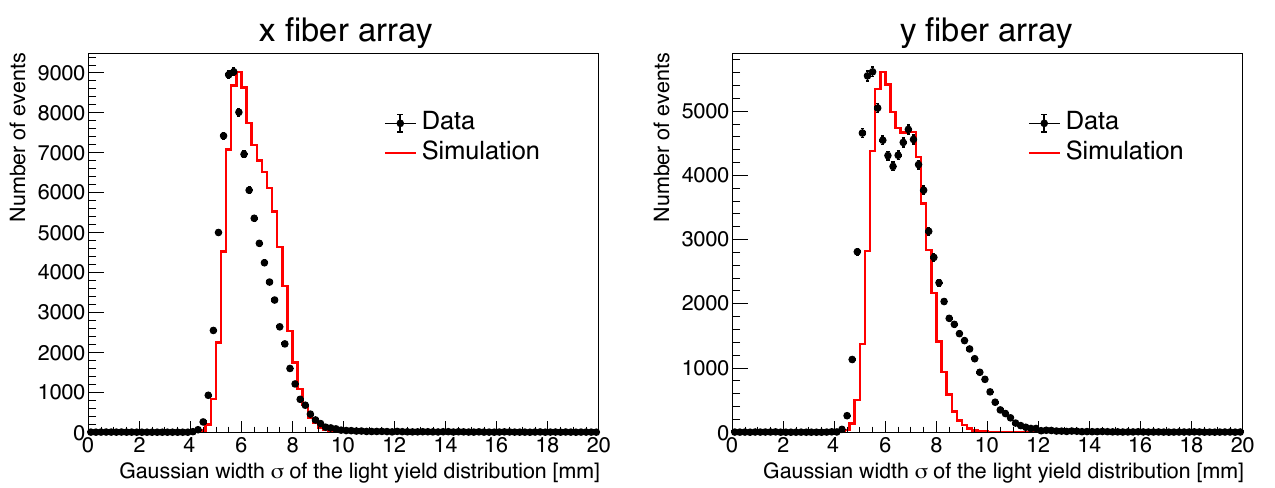}
    \caption{Data--simulation comparison of the Gaussian width $\sigma$ in the $x$ and $y$ fiber arrays for Prototype~M3 after parameter tuning. The simulation histograms are normalized such that their maximum bin contents match those of the data.}
    \label{fig:beamtest:tuning_sigma_comparison}
  \end{subfigure}
  \caption{Example of a Gaussian fit to the light yield distribution and comparison of the distribution of the fitted Gaussian width $\sigma$ between data and the tuned simulation for Prototype~M3.}
  \label{fig:beamtest:tuning_lambda}
\end{figure}

First, $\lambda_{\mathrm{scat}}$ was determined by comparing the transverse spread of the light yield distribution across the fiber channels in data and simulation.
The light yield distribution was fitted with a Gaussian function for each event, and the fitted width $\sigma$ was used as a measure of the light localization (an example for Prototype~M3 is shown in Fig.~\ref{fig:beamtest:tuning_lambda}(\subref{fig:beamtest:tuning_gauss})).
For each prototype, $\lambda_{\mathrm{scat}}$ was tuned such that the simulation reproduced the mean of the $\sigma$ distribution observed in data.
This tuning was performed independently for the $x$ and $y$ fiber arrays, and the final $\lambda_{\mathrm{scat}}$ value quoted for each prototype was taken as the average of the two results.
As a representative example, Fig.~\ref{fig:beamtest:tuning_lambda}(\subref{fig:beamtest:tuning_sigma_comparison}) compares the $\sigma$ distributions for Prototype~M3 in the $x$ and $y$ fiber arrays between data and the tuned simulation.
We note that, in Fig.~\ref{fig:beamtest:tuning_lambda}(\subref{fig:beamtest:tuning_sigma_comparison}), the data distributions in the $x$ and $y$ fiber arrays are shifted in opposite directions relative to the simulation.
This is because the $x$- and $y$-based tunings yielded slightly different $\lambda_{\mathrm{scat}}$ values (smaller in $x$ and larger in $y$), while the tuned simulation used their average.
The tuned values are summarized in Tab.~\ref{tab:beamtest:tunedparameters}.
As expected, the tuned $\lambda_{\mathrm{scat}}$ decreases with increasing scatterer concentration.

\begin{figure}[htbp]
  \centering
  \begin{subfigure}{0.5\textwidth}
    \centering
    \includegraphics[width=\linewidth]{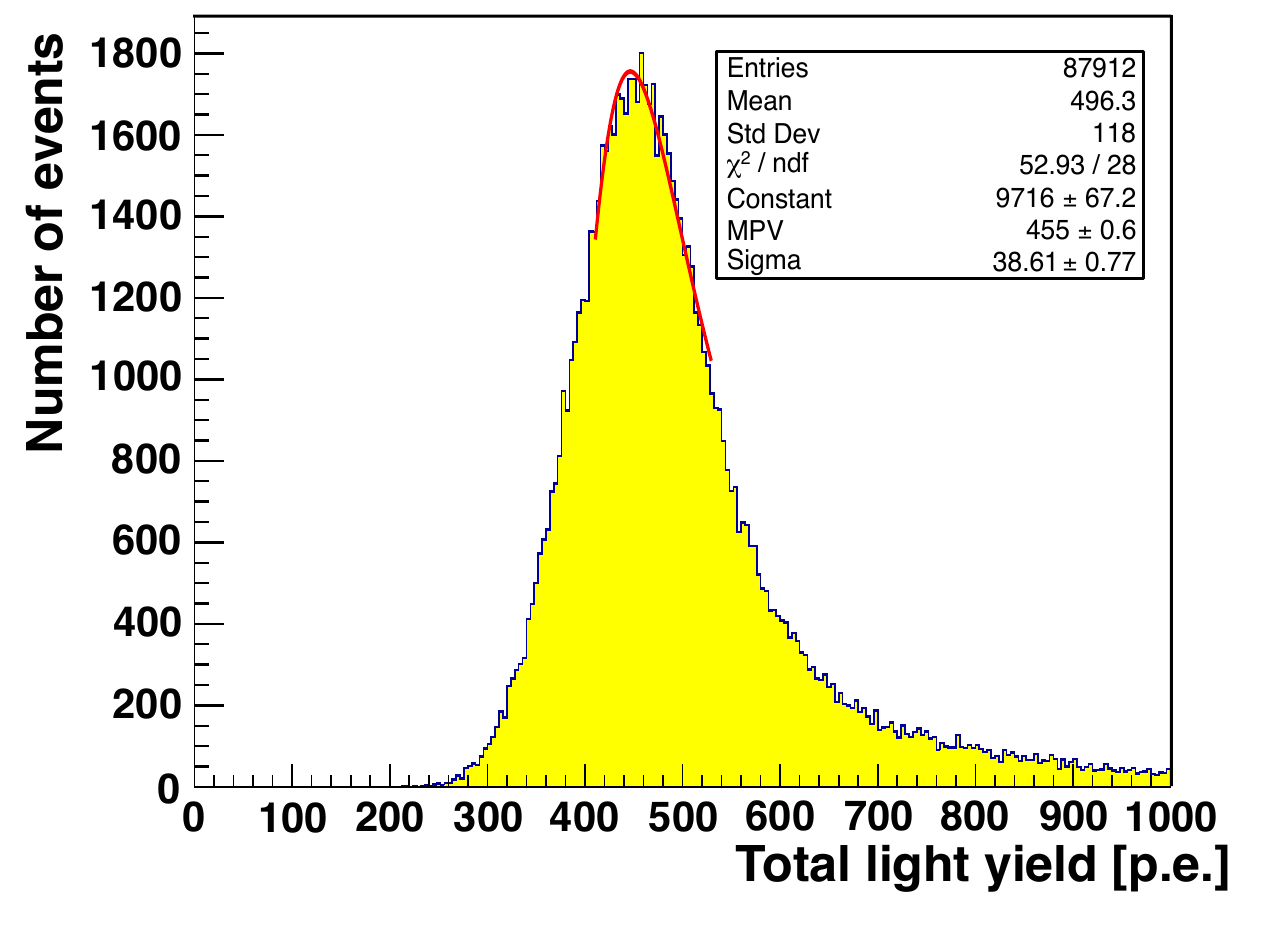}
    \caption{Example of a Landau fit to the total light yield distribution (Prototype~M3).}
    \label{fig:beamtest:tuning_landau}
  \end{subfigure}
  \hfill
  \begin{subfigure}{0.47\textwidth}
    \centering
    \includegraphics[width=\linewidth]{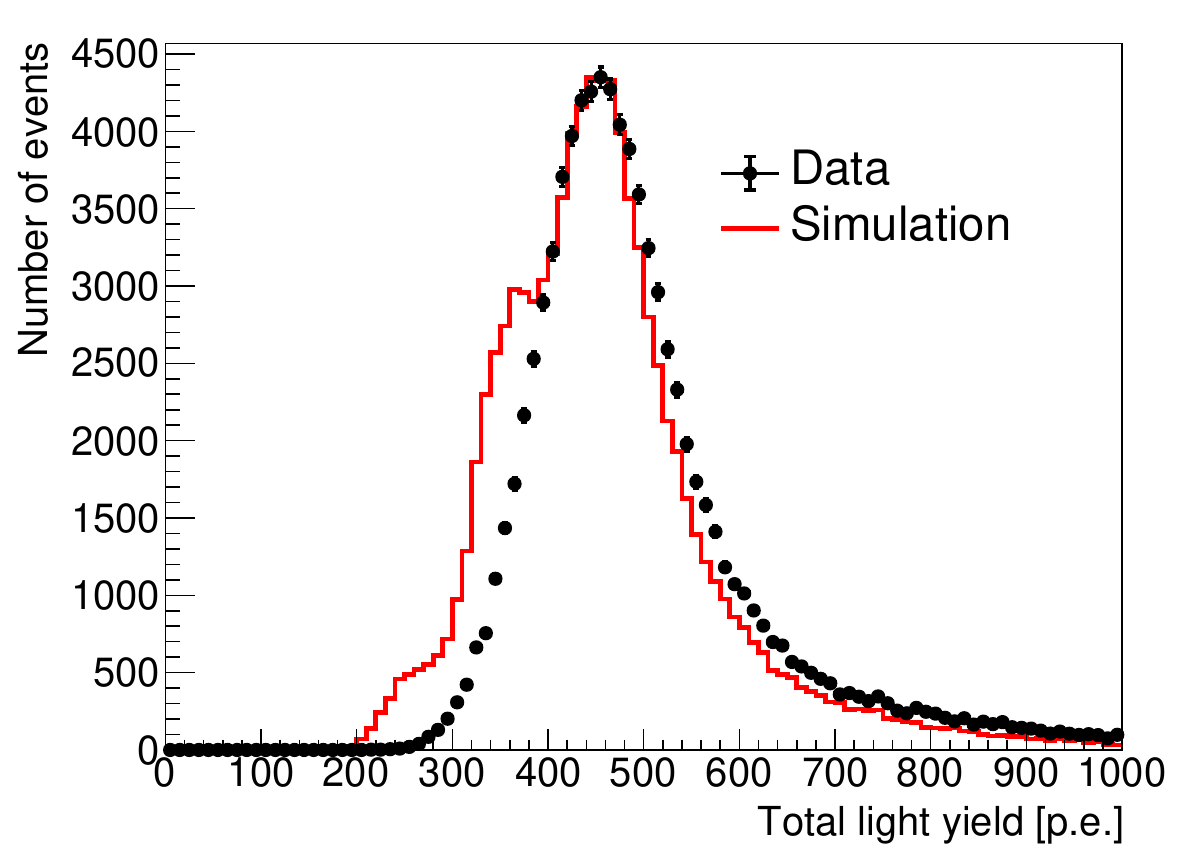}
    \caption{Data--simulation comparison of the total light yield distribution for Prototype~M3 after parameter tuning. The simulation histogram is normalized such that its maximum bin content matches that of the data.}
    \label{fig:beamtest:tuning_scintilight_comparison}
  \end{subfigure}
  \caption{Example of a Landau fit to the total light yield distribution and comparison of the total light yield distribution between data and the tuned simulation for Prototype~M3.}
  \label{fig:beamtest:tuning_scintilight}
\end{figure}

\begin{table}[htbp]
  \centering
  \caption{Tuned values of the scattering length $\lambda_{\mathrm{scat}}$, the scintillation photon yield $Y_\mathrm{scint}$, and the reflectivity scaling factor $\alpha$ for Prototypes~M1--M3. The values in parentheses in the $Y_\mathrm{scint}$ column indicate the MPV of the measured total light yield used for the $Y_\mathrm{scint}$ tuning. The uncertainty on $\lambda_{\mathrm{scat}}$ is taken as the standard deviation between the estimates obtained from the $x$ and $y$ fiber arrays.}
  \label{tab:beamtest:tunedparameters}
  \begin{tabular}{c | c c c} \hline\hline
  Prototype & $\lambda_{\mathrm{scat}}$~[mm] & $Y_\mathrm{scint}$~[photons/MeV] (MPV~[p.e.]) & $\alpha$ \\\hline
  M1 & $2.08\pm0.23$ & 6132 ($513.1\pm0.7$) & 1.0375\\
  M2 & $1.29\pm0.04$ & 5625 ($478.3\pm0.6$) & 1.0375\\
  M3 & $0.92\pm0.17$ & 5376 ($455.0\pm0.6$) & 1.0375\\\hline\hline
 \end{tabular}
\end{table}

Second, $Y_\mathrm{scint}$ was tuned by matching the total light yield, defined as the sum over all channels in both the $x$ and $y$ fiber arrays.
We used the distribution of the total light yield and extracted its most-probable value (MPV) with a Landau fit (Fig.~\ref{fig:beamtest:tuning_scintilight}(\subref{fig:beamtest:tuning_landau}) shows an example for Prototype~M3).
$Y_\mathrm{scint}$ in the simulation was adjusted so that the MPV of the total light yield matched between data and simulation.
As a representative example, Fig.~\ref{fig:beamtest:tuning_scintilight}(\subref{fig:beamtest:tuning_scintilight_comparison}) compares the total light yield distribution for Prototype~M3 between data and the tuned simulation.
It should be noted that the data and simulation show a discrepancy at low total light yield, but this has a negligible impact on the position reconstruction, which relies on the channel-to-channel light yield distribution rather than the overall light yield scale.
The resulting tuned $Y_\mathrm{scint}$ values are listed in Tab.~\ref{tab:beamtest:tunedparameters}, together with the MPV of the measured total light yield used for the tuning.
The total light yield decreases with increasing scatterer concentration, indicating absorption of scintillation photons by the scatterers.
For Prototype~T2, which had the same scatterer concentration as Prototype~M2, the tuning results obtained with Prototype~M2 data were applied to Prototype~T2 for both $\lambda_{\mathrm{scat}}$ and the scintillation photon yield.

\begin{figure}[t]
  \centering
  \begin{subfigure}{0.5\textwidth}
    \centering
    \includegraphics[width=\linewidth]{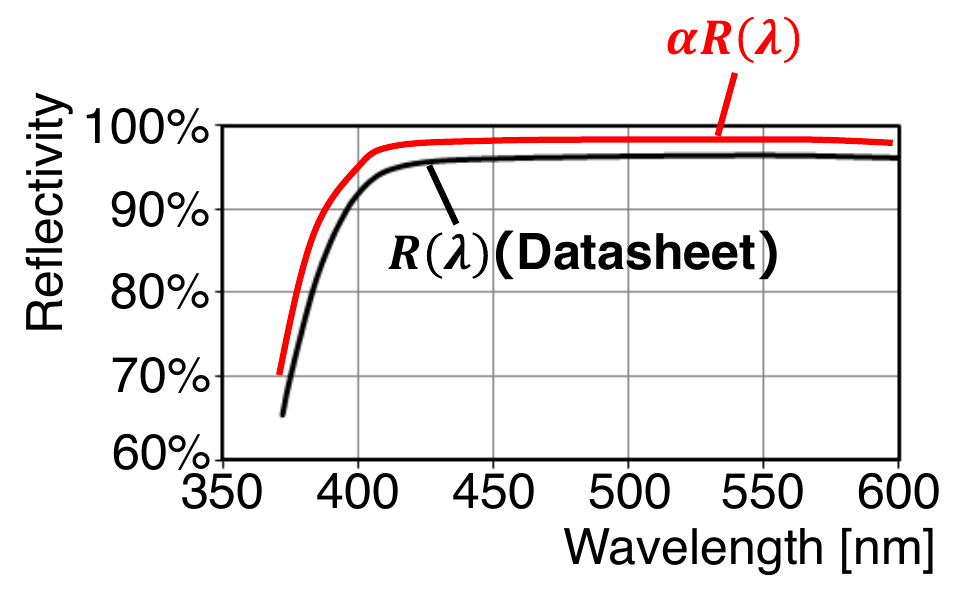}
    \caption{Reflectivity of the reflective coating~\cite{eljen:EJ510}.}
    \label{fig:beamtest:reflectivitytuning_reflectivity}
  \end{subfigure}
  \hfill
  \begin{subfigure}{0.9\textwidth}
    \centering
    \includegraphics[width=\linewidth]{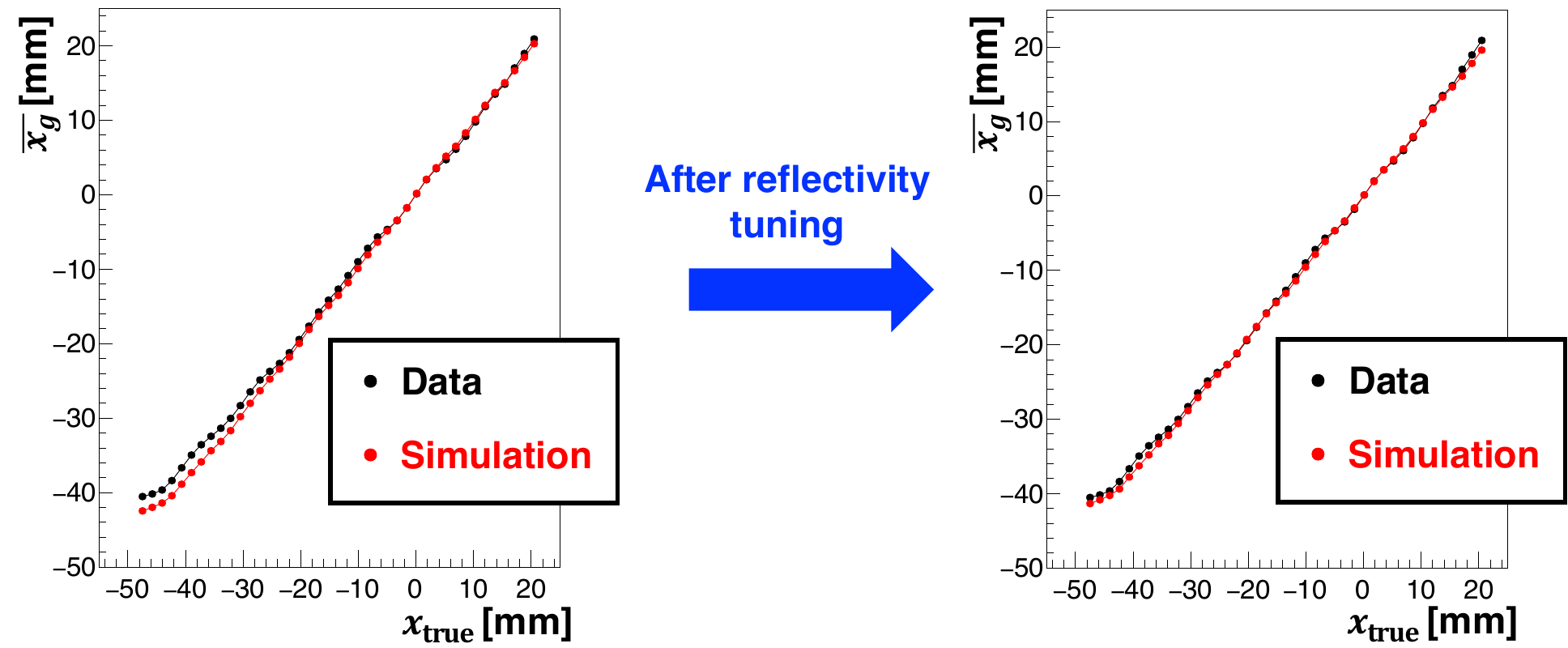}
    \caption{Average weighted center of light yield before and after tuning (Prototype~M3). Here, ``before'' denotes the simulation after tuning $\lambda_{\mathrm{scat}}$ and $Y_{\mathrm{scint}}$ but before tuning $\alpha$.}
    \label{fig:beamtest:reflectivitytuning_xg}
  \end{subfigure}
  \caption{Reflectivity of the reflective coating and comparison of the average weighted center of light yield before and after tuning for Prototype~M3.}
  \label{fig:beamtest:reflectivitytuning}
\end{figure}

Finally, the reflectivity of the reflective coating was tuned to improve the agreement between data and simulation for the weighted center of light yield.
In the simulation, the wavelength-dependent datasheet reflectivity $R(\lambda)$~\cite{eljen:EJ510} was scaled by a factor $\alpha$ (Fig.~\ref{fig:beamtest:reflectivitytuning}(\subref{fig:beamtest:reflectivitytuning_reflectivity})), and $\alpha$ was chosen to minimize the data--simulation differences in the mean weighted center of light yield
along the hodoscope-defined central cell row ($y\simeq 0~\mathrm{mm}$) and column ($x\simeq 0~\mathrm{mm}$).
This procedure was applied to the normal-incidence data for Prototypes~M1--M3, and the average value $\alpha=1.0375$ was adopted for the reflectivity scaling used throughout the simulation, including Prototype~T2 (Tab.~\ref{tab:beamtest:tunedparameters}).
As shown in Fig.~\ref{fig:beamtest:reflectivitytuning}(\subref{fig:beamtest:reflectivitytuning_xg}), this tuning improved the agreement of the weighted center of light yield response.

As an additional consideration, the attenuation length of the WLS fibers used in the prototypes was approximately 4~m, much longer than the fiber length of 130~mm. Its effect is therefore small for the beam test setup, and the datasheet value was used in the simulation without further tuning. The attenuation in the WLS fibers mainly changes the overall scale of the light yield rather than the shape of the distribution of light yield among channels, so its impact on the position reconstruction is limited. For detectors with larger areas, however, the attenuation length of the WLS fibers may need to be included as a tuning parameter to reproduce the position dependence of the total light yield in the simulation.

\subsubsection{Position reconstruction in the beam test}
\label{beamtest:posrec}

For each selected event, the weighted centers of light yield $x_g$ and $y_g$ were calculated from the measured light yields in the $x$- and $y$-fiber arrays using Eq.~\eqref{eq:xgdef}.
The reconstructed positions were then obtained as $x_{\mathrm{rec}}=f(x_g)$ and $y_{\mathrm{rec}}=f(y_g)$, where the mapping function $f$ was derived from the tuned optical simulation described in Sec.~\ref{beamtest:simtuning}.
The mapping function was constructed using normally incident 1~GeV muons injected at known positions.
Simulated events were generated at discrete $x_{\mathrm{true}}$ positions in 1-mm steps with $y_{\mathrm{true}}$ uniformly distributed in $-5~\mathrm{mm}<y_{\mathrm{true}}<5~\mathrm{mm}$, and the average weighted center $\overline{x_g}$ was computed for each $x_{\mathrm{true}}$.
The relation between $x_{\mathrm{true}}$ and $\overline{x_g}$ was then used to build $x_{\mathrm{true}}=f(\overline{x_g})$ by linear interpolation between neighboring points.
By detector symmetry, the same mapping function was used for the $y$ coordinate.

\subsection{Results}
\label{beamtest:results}

\subsubsection{Detection efficiency}
\label{beamtest:efficiency}

We evaluated the detection efficiency of the FROST prototypes using a sample of through-going positron events selected with the upstream and downstream hodoscopes.
Events were required to satisfy the following conditions:
(i) in each of the four hodoscope layers (upstream $x/y$ and downstream $x/y$), exactly one fiber had a signal larger than 10~p.e.;
(ii) the reference crossing position on FROST determined from the hodoscopes was inside the scintillator volume; and
(iii) the hit-cell difference between the upstream and downstream hodoscopes was at most one cell in both $x$ and $y$.
These requirements suppressed noise-induced hits and ensured that the positrons traversed the scintillator.
For the selected event sample, we defined the detection efficiency as the fraction of events for which the total light yield of the FROST prototype exceeded 10~p.e.\ in both $x$ and $y$.

We evaluated the detection efficiency for all datasets (Prototypes~M1--M3, T2 for normal incidence and Prototype~M3 for $\theta_x=15^\circ$, $30^\circ$, and $45^\circ$).
In all cases, the efficiency exceeded $99.99\%$ in line with the expectation of near-100$\%$ efficiency.
For the bonded prototype~T2, the efficiency also exceeded $99.99\%$ in the position cells that include the bonded interfaces, indicating that the efficiency loss due to the internal boundaries is negligible.
For comparison, the simulation with a 0.05-mm optical cement boundary gives an efficiency of 98.57\% for exactly normal incidence in the corresponding cells, which is lower than in the data, presumably because the beam test data include tracks with a finite angular spread rather than perfectly normal incidence.

\subsubsection{Evaluation of position resolution}
\label{beamtest:posres}

We evaluated the position resolution using the residuals
$x_{\mathrm{rec}}-x_{\mathrm{true}}$ and $y_{\mathrm{rec}}-y_{\mathrm{true}}$,
where $(x_{\mathrm{true}},y_{\mathrm{true}})$ was the reference crossing position determined by the upstream hodoscope.
To avoid the degradation in position reconstruction near the scintillator edges, we evaluated the resolution within the fiducial region $-40~\mathrm{mm}<x_{\mathrm{true}}<0~\mathrm{mm}$ and $0~\mathrm{mm}<y_{\mathrm{true}}<40~\mathrm{mm}$, which excluded the outermost 10~mm from the edges.
For $\theta_x=15^\circ$ and $30^\circ$, the fiducial range in $y$ was limited to $0~\mathrm{mm}<y_{\mathrm{true}}<15~\mathrm{mm}$ due to the available scan coverage.
To avoid biases from non-uniform statistics, the event counts were equalized across the selected position cells.

\begin{figure}[htbp]
  \centering
  \begin{subfigure}{0.49\textwidth}
    \centering
    \includegraphics[width=\linewidth]{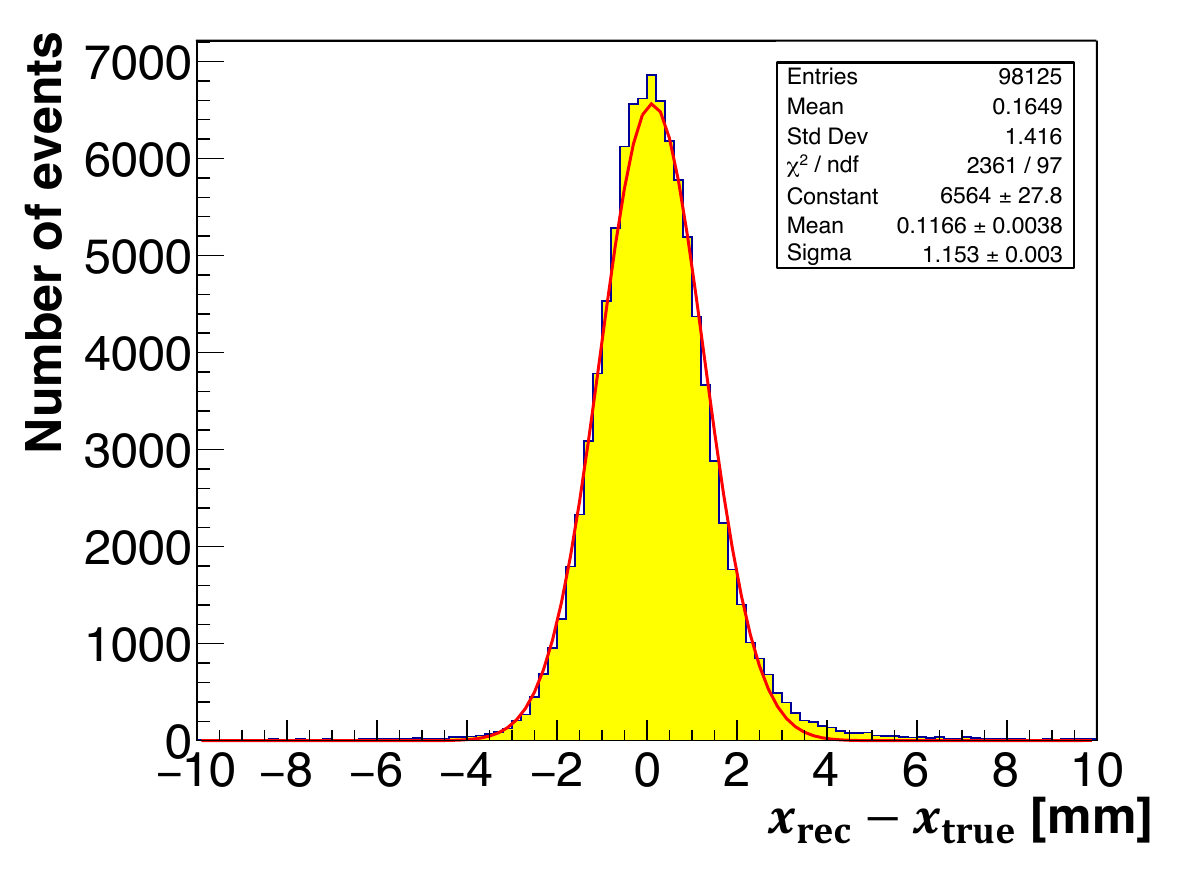}
    \caption{$\theta_x=0^\circ$}
    \label{fig:beamtest:6bai_resolution}
  \end{subfigure}
  \hfill
  \begin{subfigure}{0.49\textwidth}
    \centering
    \includegraphics[width=\linewidth]{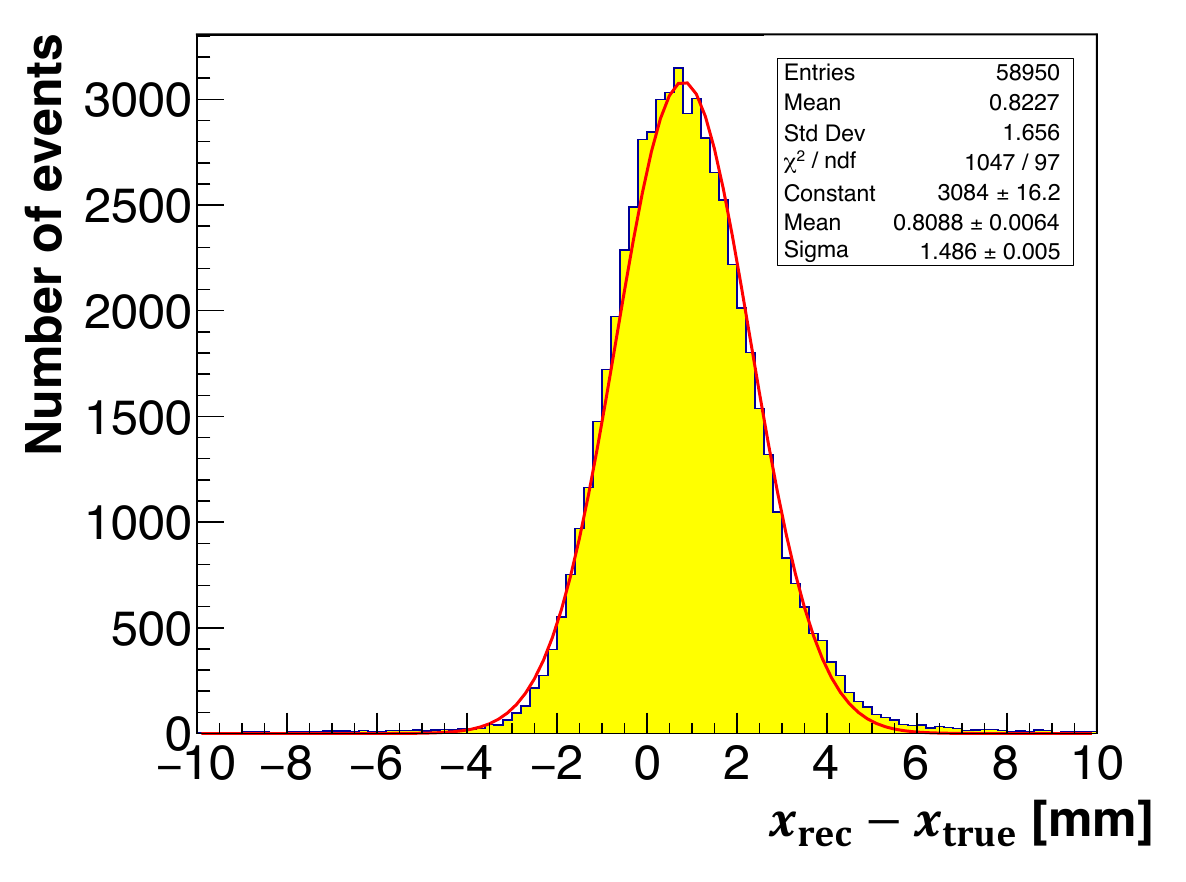}
    \caption{$\theta_x=45^\circ$}
    \label{fig:beamtest:6bai45deg_resolution}
  \end{subfigure}
  \caption{Residual distribution $x_{\mathrm{rec}}-x_{\mathrm{true}}$ for Prototype~M3 in the fiducial region.}
  \label{fig:beamtest:resolutiongauss}
\end{figure}

For each dataset, the residual distribution in the fiducial region was fitted with a Gaussian function.
We defined the Gaussian mean as $\Delta_{\mathrm{bias}}$ and the width as $\sigma_{\mathrm{Gauss}}$.
The intrinsic contribution from the FROST prototype was evaluated by subtracting the hodoscope position resolution $\sigma_{\mathrm{hodo}}$,
\begin{equation}
\sigma_{\mathrm{stat,e^+}} \equiv \sqrt{\sigma_{\mathrm{Gauss}}^2-\sigma_{\mathrm{hodo}}^2}.
\end{equation}
Assuming a uniform hit distribution within a $1.7\times1.7~\mathrm{mm}^2$ cell, we took $\sigma_{\mathrm{hodo}}=1.7~\mathrm{mm}/(\sqrt{12}\cos\theta_x)$ for $x$ and $\sigma_{\mathrm{hodo}}=1.7~\mathrm{mm}/\sqrt{12}$ for $y$.
$\sigma_{\mathrm{stat,e^+}}$ represents primarily the statistical contribution to the position resolution arising from statistical fluctuations of the detected light yield, while $\Delta_{\mathrm{bias}}$ represents primarily the systematic contribution.
Figure~\ref{fig:beamtest:resolutiongauss} shows the residual distributions for Prototype~M3 at $\theta_x=0^\circ$ and $45^\circ$.

In many applications of scintillator trackers, charged particles traverse the detector as minimum-ionizing particles (MIPs).
Since $\sigma_{\mathrm{stat,e^+}}$ scales approximately with the detected light yield to the power of $-1/2$, and the detected light yield is approximately proportional to the energy loss $\mathrm{d}E/\mathrm{d}x$, we expect $\sigma_{\mathrm{stat,e^+}}\propto 1/\sqrt{\mathrm{d}E/\mathrm{d}x}$.
We therefore converted $\sigma_{\mathrm{stat,e^+}}$ measured with the 730~MeV positron beam to the expectation for MIP-like tracks in a plastic scintillator as
\begin{equation}
\sigma_{\mathrm{stat,MIP}} = \sigma_{\mathrm{stat,e^+}} \sqrt{\frac{\left(\mathrm{d}E/\mathrm{d}x\right)_{e^+}}{\left(\mathrm{d}E/\mathrm{d}x\right)_{\mathrm{MIP}}}}
=1.08\sigma_{\mathrm{stat,e^+}} .
\end{equation}

We estimated the systematic component using $\Delta_{\mathrm{bias}}$.
For inclined incidence, a systematic shift arises because the weighted center of light yield depends on the incidence angle, while the mapping function $x_{\mathrm{true}}=f(\overline{x_g})$ is derived from normal-incidence simulation.
We therefore defined $\sigma_{\mathrm{ang}}$ as the magnitude of the simulated $\Delta_{\mathrm{bias}}$ in $x$ for $\theta_x=15^\circ$, $30^\circ$, and $45^\circ$ (and $\sigma_{\mathrm{ang}}=0$ otherwise).
The remaining contributions, such as residual relative misalignment between the hodoscopes and the FROST prototype and non-uniform optical response (e.g., scatterer/reflector non-uniformities and WLS fiber bending), were conservatively parameterized as $\sigma_{\mathrm{other}}=0.90~\mathrm{mm}$, set to the maximum magnitude of $\Delta_{\mathrm{bias}}$ observed across all measurement conditions.
The overall position resolution for MIP-like tracks was then evaluated as
\begin{equation}
\sigma_{\mathrm{pos,MIP}}=\sqrt{\sigma_{\mathrm{stat,MIP}}^2+\sigma_{\mathrm{ang}}^2+\sigma_{\mathrm{other}}^2}.
\end{equation}
Table~\ref{tab:beamtest:resolutionresult} summarizes $\sigma_{\mathrm{pos,MIP}}$ and its breakdown.

\begin{table}[htbp]
  \caption{Position resolution for MIP-like tracks and its breakdown. The resolution components are defined in Sec.~\ref{beamtest:posres}.}
  \label{tab:beamtest:resolutionresult}
  \centering
  \begin{subtable}{1.0\textwidth}
  \centering
  \caption{$x$ direction}
  \label{tab:beamtest:resolutionx}
  {\small
  \begin{tabular}{c|c|c|c|c|c} \hline\hline
  \multicolumn{2}{c|}{}  & \multicolumn{3}{c|}{Breakdown} & Position resolution \\ \hline
  Prototype & $\theta_x$ & $\sigma_{\mathrm{stat,MIP}}$~[mm] &$\sigma_{\mathrm{ang}}$~[mm] & $\sigma_{\mathrm{other}}$~[mm] & $\sigma_{\mathrm{pos,MIP}}$~[mm]\\ \hline
  M1 &  & $1.34$ & \multirow{4}{*}{$0$} & \multirow{7}{*}{$0.90$} & $1.61$ \\ \cline{1-1}\cline{3-3}\cline{6-6}
  M2 & $0^\circ$ & $1.13$ &  &  & $1.44$ \\ \cline{1-1}\cline{3-3}\cline{6-6}
  M3 &  & $1.12$ &  &  & $1.44$ \\ \cline{1-1}\cline{3-3}\cline{6-6}
  T2 &  & $1.12$ &  &  & $1.44$ \\ \cline{1-2}\cline{3-4}\cline{6-6}
   & $15^\circ$ & $1.20$ & $0.20$ &  & $1.51$ \\ \cline{2-2}\cline{3-4}\cline{6-6}
  M3 & $30^\circ$ & $1.22$ & $0.46$ &  & $1.58$ \\ \cline{2-2}\cline{3-4}\cline{6-6}
   & $45^\circ$ & $1.41$ & $0.79$ &  & $1.85$ \\ \hline\hline
  \end{tabular}
  }
  \end{subtable}
  \hfill
  \vspace{0.3cm}
  \begin{subtable}{1.0\textwidth}
  \centering
  \caption{$y$ direction}
  \label{tab:beamtest:resolutiony}
  {\small
  \begin{tabular}{c|c|c|c|c|c} \hline\hline
  \multicolumn{2}{c|}{}  & \multicolumn{3}{c|}{Breakdown} & Position resolution \\\hline
  Prototype & $\theta_x$ & $\sigma_{\mathrm{stat,MIP}}$~[mm] &$\sigma_{\mathrm{ang}}$~[mm] & $\sigma_{\mathrm{other}}$~[mm] & $\sigma_{\mathrm{pos,MIP}}$~[mm]\\ \hline
  M1 &  & $1.54$ & \multirow{7}{*}{$0$} & \multirow{7}{*}{$0.90$} & $1.78$ \\ \cline{1-1}\cline{3-3}\cline{6-6}
  M2 & $0^\circ$ & $1.31$ &  &  & $1.59$ \\ \cline{1-1}\cline{3-3}\cline{6-6}
  M3 &  & $1.19$ &  &  & $1.49$ \\ \cline{1-1}\cline{3-3}\cline{6-6}
  T2 &  & $1.36$ &  &  & $1.63$ \\ \cline{1-2}\cline{3-3}\cline{6-6}
    & $15^\circ$ & $0.96$ &  &  & $1.32$ \\ \cline{2-2}\cline{3-3}\cline{6-6}
  M3 & $30^\circ$ & $0.93$ &  &  & $1.29$ \\ \cline{2-2}\cline{3-3}\cline{6-6}
    & $45^\circ$ & $1.09$ &  &  & $1.41$ \\ \hline\hline
  \end{tabular}
  }
  \end{subtable}
\end{table}

Comparing Prototypes~M1--M3, the position resolution improved with increasing scatterer concentration,
indicating that the benefit of stronger light localization dominated over the loss of light yield due to absorption by the scatterers for the scatterer concentrations studied here.
The scatterer concentration of Prototype~M3 provided the best balance between improved light localization and reduced light yield, and was therefore selected as the optimal choice for the present FROST design.

For quantitative comparison, we also quote the normalized position resolution $\sigma_{\mathrm{pos}}/w$, where $\sigma_{\mathrm{pos}}$ denotes the position resolution and $w$ is the readout pitch.
This dimensionless quantity quantifies performance per readout channel.
For normal incidence, Prototype~M3 provided the best performance, with an average position resolution of $1.47~\mathrm{mm}$ (mean of the $x$ and $y$ resolutions), corresponding to $\sigma_{\mathrm{pos,MIP}}/w=0.147$.
This value was well below the value of $1/\sqrt{12}=0.289$ expected for a purely segmented detector with a uniform hit probability within a segment.

For inclined incidence with Prototype~M3, the $x$-direction resolution degraded with increasing $\theta_x$.
This trend was driven by the increase of $\sigma_{\mathrm{stat,MIP}}$ because the track spanned a wider range in $x$, and by the increase of $\sigma_{\mathrm{ang}}$ due to the angle dependence of the weighted center of light yield relative to the normal-incidence mapping function.
We also note that $\sigma_{\mathrm{stat,MIP}}$ in the $y$ direction became smaller than in the normal-incidence case, which is consistent with the longer path length in the scintillator and the resulting higher light yield for inclined tracks.
The resulting MIP-equivalent resolution in $x$ was $1.51~\mathrm{mm}$, $1.58~\mathrm{mm}$, and $1.85~\mathrm{mm}$ for $\theta_x=15^\circ$, $30^\circ$, and $45^\circ$, respectively,
corresponding to $\sigma_{\mathrm{pos,MIP}}/w=0.151$, $0.158$, and $0.185$.
Even at $\theta_x=45^\circ$, $\sigma_{\mathrm{pos,MIP}}/w$ remained well below $1/\sqrt{12}=0.289$.

\begin{figure}[htbp]
  \centering
  \begin{subfigure}{0.49\textwidth}
    \centering
    \includegraphics[width=\linewidth]{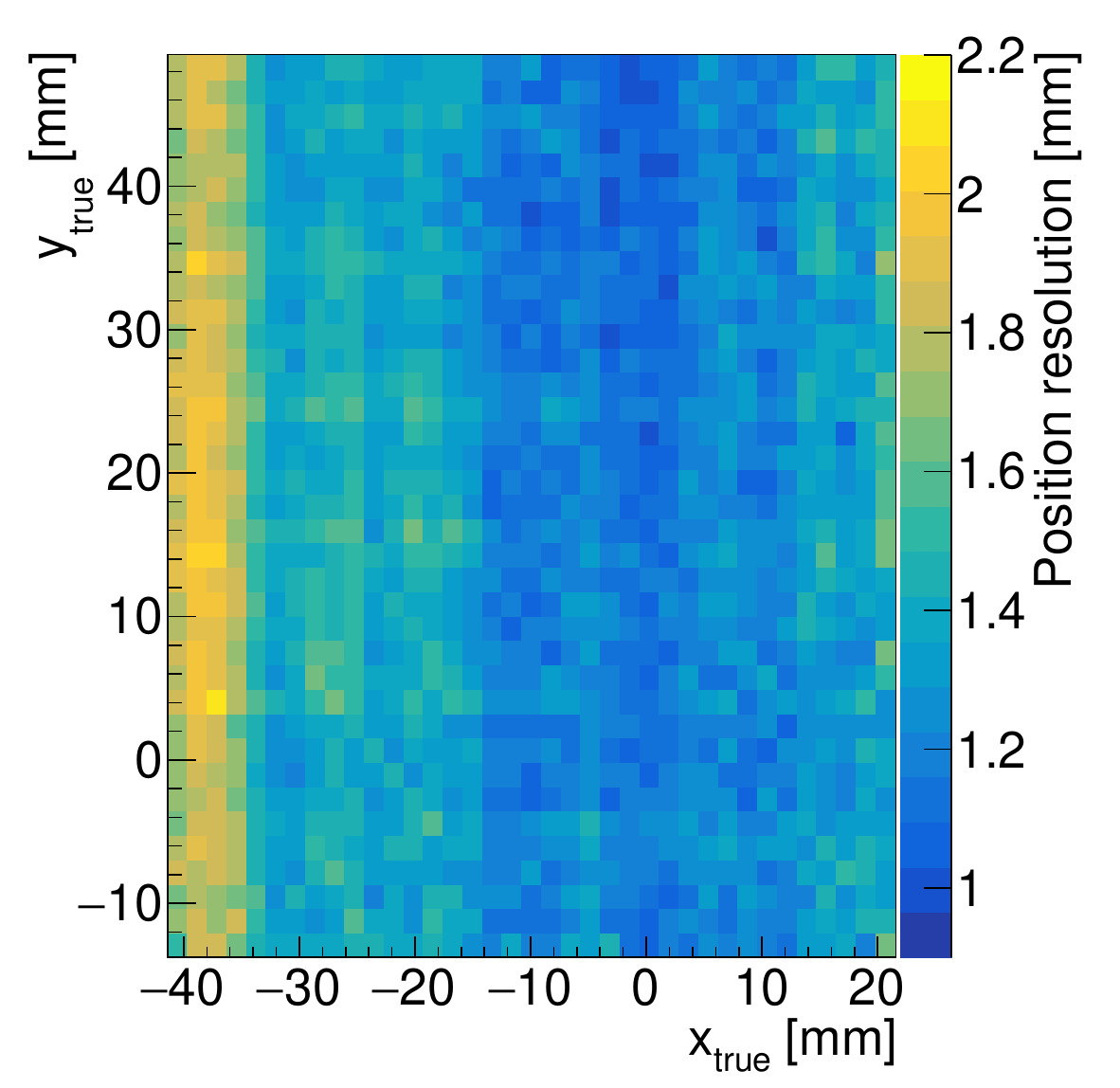}
    \caption{Prototype~M2.}
    \label{fig:beamtest:4bai_resolutionx}
  \end{subfigure}
  \hfill
  \begin{subfigure}{0.49\textwidth}
    \centering
    \includegraphics[width=\linewidth]{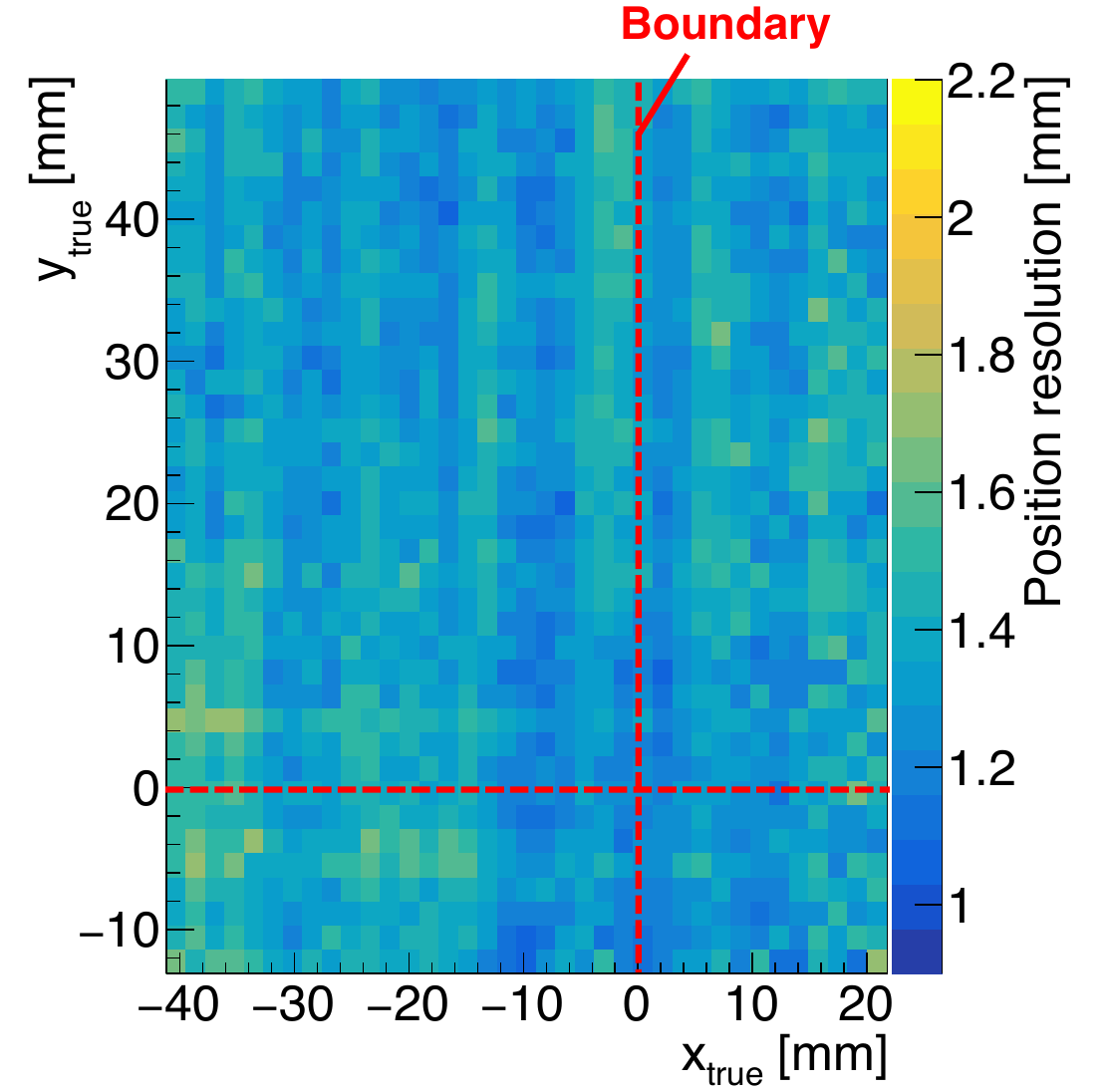}
    \caption{Prototype~T2.}
    \label{fig:beamtest:4baihariawase_resolutionx}
  \end{subfigure}
  \caption{Cell-by-cell $x$ position resolution for Prototypes~M2 and~T2 in the region $x_{\mathrm{true}}>-40~\mathrm{mm}$.}
  \label{fig:beamtest:hariawase_resolutionx}
\end{figure}

Finally, comparing Prototypes~M2 and~T2, which shared the same scatterer concentration, the position resolution agreed within $3\%$ in the fiducial region.
Figure~\ref{fig:beamtest:hariawase_resolutionx} compares the cell-by-cell $x$ position resolution for Prototypes~M2 and~T2 in the region $x_{\mathrm{true}}>-40~\mathrm{mm}$.
The differences between the two prototypes were small compared to the cell-to-cell variations within each prototype.
The cell-to-cell variations were attributed to the same effects parameterized by $\sigma_{\mathrm{other}}$.
Together with the result that the efficiency loss due to the internal boundaries was negligible in Prototype~T2 (Sec.~\ref{beamtest:efficiency}), these results indicate that assembling a monolithic-equivalent plate
by bonding scintillator tiles with optical cement is a viable approach for scaling FROST to larger areas without degrading the detector performance.


\section{Conclusion}
\label{sec:conclusion}

In this paper, we proposed FROST (Fiber-Readout mOnolithic and Scatterer\hyp embedded scintillator Tracker) as a new scintillator tracker concept.
We demonstrated its principle and evaluated its performance using prototype detectors in the positron beam test.
FROST is based on a monolithic plastic scintillator plate with embedded scatterers, with scintillation light collected by WLS fibers in surface grooves and read out by SiPMs.
Embedded scatterers localize scintillation light, and channels closer to the charged-particle crossing point have larger detected light yield.
The position is reconstructed based on the light yield distribution across channels.
The positron beam test at RARiS was performed with FROST prototypes featuring different specifications to evaluate the detection efficiency and position resolution for different scatterer concentrations, and to study the impact of scintillator bonding.
The results confirmed that the position-reconstruction principle worked well.
The efficiency exceeded $99.99\%$ for all configurations.
The best normal-incidence performance was achieved with the prototype with the highest scatterer concentration, giving $1.47~\mathrm{mm}$ ($\sigma_{\mathrm{pos,MIP}}/w=0.147$), and the resolution remained $1.85~\mathrm{mm}$ even for a $45^\circ$ incidence ($\sigma_{\mathrm{pos,MIP}}/w=0.185$), well below the segmented expectation ($\sigma_{\mathrm{pos,MIP}}/w=1/\sqrt{12}=0.289$).
A bonded prototype showed agreement in position resolution within $3\%$ with the monolithic case, and no measurable efficiency loss was observed at the bonded interfaces, indicating that optical-cement bonding can scale FROST while preserving the position resolution.
In summary, these results demonstrated the feasibility of FROST as a scintillator tracker achieving a position resolution well below the 10-mm readout pitch.

\section*{Declaration of generative AI and AI-assisted technologies in the manuscript preparation process}

During the preparation of this work, the authors used ChatGPT in order to improve the readability and language of the manuscript. After using this tool, the authors reviewed and edited the content as needed and take full responsibility for the content of the published article.

\section*{CRediT authorship contribution statement}

\textbf{Naoki Otani}: Formal analysis, Funding acquisition, Investigation, Methodology, Software, Validation, Visualization, Writing -- original draft, Writing -- review \& editing. \textbf{Seungho Han}: Investigation, Writing -- review \& editing. \textbf{Shun Ito}: Investigation, Writing -- review \& editing. \textbf{Tatsuya Kikawa}: Conceptualization, Investigation, Methodology, Supervision, Writing -- review \& editing. \textbf{Tsuyoshi Nakaya}: Conceptualization, Funding acquisition, Investigation, Methodology, Supervision, Writing -- review \& editing. \textbf{Mihiro Suzuki}: Investigation, Writing -- review \& editing. \textbf{Atsushi Tokiyasu}: Investigation, Resources, Writing -- review \& editing.

\section*{Acknowledgments}
Part of this study was performed using facilities of the Research Center for Accelerator and Radioisotope Science (RARiS), Tohoku University (Proposal No. 3052).
The authors thank RARiS for the allocation of beamtime and the RARiS staff for their kind support during the beam test.
We thank Haruka Komaba for her valuable support during the beam test.
We are grateful to Kuraray for accommodating our requests regarding the scatterer-embedded scintillator and for fabricating the scintillator plates used in the FROST prototypes.
We also thank Shigeki Aoki and Atsumu Suzuki for sharing their experience with detectors based on scatterer-embedded scintillators and for valuable advice.

This work was supported by JSPS KAKENHI Grant Numbers JP23H05434, JP25KJ1470.

\end{document}